# The emergence, coalescence and topological properties of multiple exceptional points and their experimental realization

Kun Ding[†], Guancong Ma[†], Meng Xiao, Z. Q. Zhang, and C. T. Chan[★]

*Department of Physics and Institute for Advanced Study,
Hong Kong University of Science and Technology,
Clear Water Bay, Kowloon, Hong Kong*

[†]These authors contributed equally.
[★]Corresponding E-mail: phchan@ust.hk

Non-Hermitian systems distinguish themselves from Hermitian systems by exhibiting a phase transition point called an exceptional point (EP), which is the point at which two eigenstates coalesce under a system parameter variation. Many interesting EP phenomena such as level crossings/repulsions in nuclear/molecular and condensed matter physics, and unusual phenomena in optics such as loss-induced lasing and unidirectional transmission can be understood by considering a simple 2x2 non-Hermitian matrix. At a higher dimension, more complex EP physics not found in two-state systems arises. We consider the emergence and interaction of multiple EPs in a four-state system theoretically and realize the system experimentally using four coupled acoustic cavities with asymmetric losses. We find that multiple EPs can emerge and as the system parameters vary, these EPs can collide and merge, leading to higher order singularities and topological characteristics much richer than those seen in two-state systems.

Non-Hermitian systems[1-3] such as open and/or lossy systems are ubiquitous in nature. Systems with parity-time symmetry[4], as a subset of non-Hermitian systems, have generated great interest lately due to a rich array of novel phenomena including a divergent Petermann factor[5,6], loss-induced revival of lasing[7], single-mode lasers[8,9], reversed pump dependence of lasers[10], Bloch oscillation[11], coherent absorption[12], optical isolation[13], unidirectional light propagation[14,15] and others[16-22]. Many of these novel phenomena can be traced to the existence of an "exceptional point" (EP) when two quasi-bound states coalesce, which is perhaps the single most important characteristic of non-Hermitian physics. The EP can be described locally by a non-diagonalizable $2\times 2$ matrix in which the eigenvalues have a square-root singularity and the eigenstates exhibit peculiar topological properties[23,24]. Periodic systems can support more complex phenomena such as a ring of EPs, but even these complex configurations can be considered using a 2x2 matrix[25]. However, non-Hermitian systems exist that cannot be described by 2x2 matrices. In those multi-state systems, multiple EPs can form[26-28] and their interactions may lead to new physics including the coalescence of two or more EPs, and new singularities with different topological properties. In this work, we investigate a four-state system both theoretically and experimentally. The emergence of multiple EPs, their topological properties, and their coalescence can be best summarized in a phase diagram featuring an exceptional point formation pattern (EPFP). The coalescence of two EPs and that of three EPs form two curves in the parameter space, partitioning the phase space into three regions each with a unique EPFP. Together with a two-state inversion line, the phase space is further divided into five regions, each with distinct topological properties. The coalescence of three EPs produces a higher order singularity and the coalescence of a pair of EPs with the same chirality produces a linear crossing that is qualitatively different from the crossing at a diabolic point.

**Two coupled acoustic cavity resonators**

We begin with a two-state system containing two coupled cavities A and B having the same resonant frequency $\omega_2$ as shown in the inset of Fig. 1a. The Hamiltonian of the system can be written as

$$H = \begin{pmatrix} \omega_2 - i\Gamma_0 & \kappa \\ \kappa & \omega_2 - i\Gamma \end{pmatrix}, \qquad (1)$$

where $\kappa$ denotes the strength of the coupling, $\Gamma_0$ denotes the intrinsic loss of each cavity and $\Gamma = \Gamma_0 + \Delta\Gamma$ with $\Delta\Gamma$ representing an additional tunable loss introduced at cavity B. The eigenfrequencies of equation (1) take the form

$$\tilde{\omega}_j = \omega_2 - i\frac{\Gamma + \Gamma_0}{2} \pm \frac{1}{2}\sqrt{4\kappa^2 - (\Delta\Gamma)^2} . \qquad (2)$$

When $\Delta\Gamma$ is increased and becomes 2|κ|, this two-cavity system will exhibit an EP at the square-root branch point. At this point, the two eigenstates coalesce and one becomes defective. Beyond this point, the imaginary part of the frequency (the "width") of the two states bifurcates. A realization of the Hamiltonian (equation (1)) using two couple acoustic cavities is shown in Fig. 1a, with details given in the Methods section. The existence of an acoustic EP in such a system is experimentally demonstrated in Figs. 1b and 1c.

**Four-state non-Hermitian Hamiltonian with coupling**

Using the above two-state system as the building block, we now construct a four-state system as shown in the inset of Fig. 2a. The system consists of two pairs of coupled cavities with the same values of $\kappa$, $\Gamma_0$ and $\Gamma$ but different resonant frequencies.

Cavities A and B form one pair with resonant frequency $\omega_2$ and cavities C and D form another pair with resonant frequency $\omega_1$. Coupling between these two pairs is introduced by connecting cavities A and D with a small tube and cavities B and C with another small tube as shown in the inset of Fig. 2a. The Hamiltonian of the system can be written as

$$H = \begin{pmatrix} \omega_2 - i\Gamma_0 & \kappa & 0 & t \\ \kappa & \omega_2 - i\Gamma & t & 0 \\ 0 & t & \omega_1 - i\Gamma_0 & \kappa \\ t & 0 & \kappa & \omega_1 - i\Gamma \end{pmatrix}, \quad (3)$$

where $t$ denotes the strength of inter-pair coupling. The eigenfrequencies of equation (3) take the following form (see Supplementary Section 1):

$$\tilde{\omega}_j = \omega_0 - i\frac{\Gamma + \Gamma_0}{2} \pm \frac{1}{2}\sqrt{\Delta_1 \pm 4\sqrt{\Delta_2}}, \quad j = 1, 2, 3, 4, \quad (4)$$

where $\omega_0 = (\omega_1 + \omega_2)/2$ and

$$\Delta_1 = -(\Delta\Gamma)^2 + 4\kappa^2 + 4t^2 + (\Delta\omega)^2, \quad (5)$$

$$\Delta_2 = 4\kappa^2 t^2 + \kappa^2(\Delta\omega)^2 - (\Delta\Gamma)^2 \frac{(\Delta\omega)^2}{4}, \quad (6)$$

with $\Delta\omega = \omega_1 - \omega_2$. Equation (4) shows that states can coalesce under three conditions, which we refer to as CS-1 to CS-3. In CS-1, CS-2, and CS-3, $\Delta_1 \pm 4\sqrt{\Delta_2} = 0$, $\Delta_2 = 0, \Delta_1 \neq 0$, and $\Delta_1 = \Delta_2 = 0$, respectively. In CS-1, only one state is defective, which corresponds to an EP. In CS-2 and CS-3, two and three states are defective, respectively. These singularities display rich topological properties which will be analyzed later using eigenvectors.

**Eigenfrequency phase diagram**

From equations (5) and (6), we see that depending on the parameters $(\Delta\omega)^2$, $\kappa^2$ and $t^2$, different combinations of CS-1, CS-2 and CS-3 may appear in the EPFP when $\Delta\Gamma$ is increased continuously. Fig. 2b shows a phase diagram in the space of two dimensionless parameters $(\Delta\omega/2\kappa)^2$ and $(t/\kappa)^2$. Three regions exist, designated Classes I, II, and III, with their boundaries marked by a solid yellow line and a solid red line. Each class represents a distinct EPFP, in which the EPs can have different singularity types. In addition, Class II and Class III can each be further divided into two topologically distinct regions, designated *a* and *b* and separated by a white dashed line. It will be shown later that while regions *a* and *b* share the same EPFP, they exhibit different topological characteristics.

To show the EPFP in each region, we choose a value of $(t/\kappa)^2 = 0.7744$ and then increase $(\Delta\omega/2\kappa)^2$ incrementally from zero, as marked by the vertical red arrow line in Fig. 2b. At $(\Delta\omega/2\kappa)^2 = 0$, two EPs (CS-1$_-$ and CS-1$_+$ corresponding respectively to $\Delta_1 - 4\sqrt{\Delta_2} = 0$ and $\Delta_1 + 4\sqrt{\Delta_2} = 0$) will form with increasing $\Delta\Gamma$ as shown in Fig. 3a, which can be obtained from equations (4-6) (see Supplementary Section 6). The EPFP of Class III-a is shown in Fig. 3b when $(\Delta\omega/2\kappa)^2 = 0.015$. In addition to the two EPs shown in Fig. 3a, another two CS-2 singularities will appear at a larger $\Delta\Gamma_{CS-2}$ for a non-zero $\Delta\omega$. The value of $\Delta\Gamma_{CS-2}$ can be obtained from equation (6) (see Supplementary Section 6), i.e.,

$$\Delta\Gamma_{CS-2}^2 = 4\kappa^2\left(1 + \frac{4t^2}{(\Delta\omega)^2}\right). \tag{7}$$

Equation (7) shows that $\Delta\Gamma_{CS-2}$ approaches infinity as $\Delta\omega$ approaches zero and these two CS-2 singularities will gradually approach the CS-1$_+$ singularity as

$(\Delta\omega/2\kappa)^2$ increases.

When $(\Delta\omega/2\kappa)^2$ is further increased, we cross the white dashed line expressed by $4\kappa^2 = 4t^2 + (\Delta\omega)^2$, which corresponds to the degeneracy condition of the two middle states in the absence of $\Delta\Gamma$ (see Supplementary Section 3). When $(\Delta\omega/2\kappa)^2 = 0.226$, the system configuration resides at the boundary of Classes IIIa and IIIb which is marked by the white dashed line as shown in Fig. 3c. On this line, the first EP (CS-1$_-$) disappears. An EP of opposite chirality reemerges when $(\Delta\omega/2\kappa)^2$ is further increased and the eigenfrequencies of the two middle states are inverted (Fig. 3d). Although the EPFPs in Fig. 3b and Fig. 3d appear to be the same when the system parameters cross the white dashed line (and hence they are called Class III-a and Class III-b, respectively), they have different chiralities associated with the CS-1$_-$ singularity.

Upon a further increase in $(\Delta\omega/2\kappa)^2$, the EPFP will switch from Class III-b (Fig. 3d) to Class II-b (Fig. 3f) as the system parameters cross the solid red line in Fig. 2b which has the form

$$(\Delta\omega)^2 = 2\sqrt{t^4 + 4\kappa^2 t^2} - 2t^2. \tag{8}$$

Equation (8) is obtained by eliminating $\Delta\Gamma$ in equations $\Delta_1 = 0$ and $\Delta_2 = 0$ (see equations (5) and (6) and Supplementary Sections 3 and 7). As shown in Fig. 3e, the configurations on this red line always carry a CS-3 type singularity for some particular values of $\Delta\Gamma$ and 3 states are defective at the CS-3 points. Such a singularity is a higher order EP[26]. The red solid line hence represents a line consisting of high order singularities. In addition to the CS-3 singularity, there exists another CS-1$_-$ singularity at a smaller value of $\Delta\Gamma$. Below this red line the CS-3 singularity splits into three singularities: one CS-1$_+$ on the left and two CS-2 on the right having the same $\text{Re}(\tilde{\omega}_j)$ but different $\text{Im}(\tilde{\omega}_j)$. Above the red line (in the Class II region), the CS-3 singularity also splits into three singularities but in a different manner: one CS-1 on the left and two CS-2 on the right. A typical EPFP is shown in Fig. 3f, in which the two CS-1-type singularities on the left are given by the two roots of $F \equiv \Delta_1 - 4\sqrt{\Delta_2} = 0$.

Increasing $(\Delta\omega/2\kappa)^2$ further will bring the system to the yellow solid line in Fig. 2b, which is given by $\alpha = \frac{(\Delta\omega)^2}{4\kappa^2} - 1 = 0$. This line separates Class II from Class I. When this line is approached from below (Class II), the two CS-1 singularities coalesce to form a linear crossing, as shown in Fig. 3g. This line is obtained from two conditions: $F = \Delta_1 - 4\sqrt{\Delta_2} = 0$, and $\frac{\partial F}{\partial(\Delta\Gamma)} = 0$ (see Supplementary Sections 3 and 5). We emphasize here that one state is defective at the linear crossing point induced by the coalescence of two EPs of the same chirality and as such, this linear crossing point is different from the diabolic point in a Hermitian Hamiltonian. Interestingly, the yellow and red solid lines converge in the limit of large $t^2/\kappa^2$, which can be seen from equation (8). After crossing the yellow line, the system enters the Class I region, with a typical EPFP shown in Fig. 3h. In this figure, the linear crossing seen in Fig. 3g has disappeared, and level repulsion of the two middle states is observed.

Similarly, Class II-a and Class II-b share the same EPFP except that the first CS-1 singularity in Fig. 3f changes chirality in the Class II-a region. Due to this change, the coalescence of two CS-1 singularities in Class II-a produces different topological characteristics than that in Class II-b. A detailed analysis can be found in Supplementary Section 9. To sum up, this system shows three classes of EPFPs, two of which can be further divided into two subclasses due to the difference in chirality of the EPs.

**Realization of exceptional point formation patterns (EPFPs)**

The phase diagram shown in Fig. 2b can be realized using coupled acoustic resonators. A photograph of the experimental setup is shown in Fig. 2a. We connected two pairs of acoustic cavities together with another two small side tubes. Again, the inter-pair coupling $t$ is determined by the cross-sectional area of the tubes. The system is pumped incoherently in four cavities to excite all possible modes. Microphones are used to measure the pressure at cavity B (filled markers) and cavity D (open makers). Additional loss $\Delta\Gamma$ is gradually increased only in cavities B and D by adding a mixture of sponge and putty. An example of a Class I EPFP is shown in Fig. 4a, with the detailed experimental parameters given in the Methods section. It can be seen that initially there are two peaks in both cavity pairs (denoted by M and N). As $\Delta\Gamma$ increases, the states belonging to the same pair of cavities coalesce (O and P). This generates two EPs (at 3436.2 Hz and 3471.0 Hz). Due to a large $\Delta\omega$, the two pairs are well separated in the frequency spectrum, so they may be considered as forming their own EPs nearly independently. The experimental results are fitted to acquire the parameters for its Hamiltonian. These fitted results are shown by dots in Fig. 4b, where the solid curves are theoretical results. Two EPs are clearly identified, a clear signature of a Class I EPFP (grey area in Fig. 2b).

To experimentally demonstrate a Class II EPFP, we decrease the eigenfrequency difference $\Delta\omega$ by reducing the height difference between inter-pair cavities, and modify the coupling strengths $\kappa, t$ by controlling the cross section of the side tubes (detailed experimental parameters are given in the Methods section). An essential condition here is that the real part of the frequency of the middle two states is inverted in the frequency spectrum, which can be achieved with a smaller difference in eigenfrequency $\Delta\omega$, or a sufficiently strong intra-pair coupling $\kappa$, or both. As shown in Fig. 4c, we can observe four peaks when there is no additional loss (denoted by M), with the two states in the middle inverted (blue and red curves in Fig. 4d). With increasing $\Delta\Gamma$, these two states coalesce with an EP occurring and the system enters a regime with three different *Re[f]* (N). However, further increases in $\Delta\Gamma$ cause the eigenfrequencies of the upper state (green curves in Fig. 4d) and lower state (black curves in Fig. 4d) to move towards the central coalesced state at 3450 Hz, eventually reaching a point where the intra-pair interactions are strong enough to pull the central state apart. This gives rise to the second EP, and the system reverts to the previous configuration with four different *Re[f]* (O). If we focus on the evolution of the two middle states, it appears that these two states coalesce as $\Delta\Gamma$ increases but bifurcate again as $\Delta\Gamma$ increases further. This cannot occur in a 2x2 system, but is allowed in higher dimensions. Eventually, $\Delta\Gamma$ is sufficiently large for the intra-pair states to coalesce. Two more EPs are generated, and the system reaches its terminal stage with states of two different *Re[f]* (P).

Upon further decreasing $\Delta\omega$ and modifying $\kappa, t$, our system enters Class III. The results of experiments and theoretical fittings are shown in Figs. 4e and 4f in a similar manner. Following the increase in $\Delta\Gamma$, the system behaves similarly to Class II in the configuration with no additional loss: the number of peaks starts at four (M, N), and then the two middle states coalesce producing the first EP, which indicates the system has entered a stage with three different *Re[f]* (O). Subsequently, the upper and lower two states also coalesce (the second EP) and the system enters a regime with only one real frequency (P). Eventually, with a sufficiently large $\Delta\Gamma$, the system generates another two EPs and reaches the final stage with states of two different *Re[f]*

(Q).

**Topological characteristics around singularities**

It is well known that a parameter variation encircling an EP will cause the two states to switch position after one cycle and acquire a geometric phase $\pm\pi$ after two cycles[23,24]. Thus, four cycles are needed to restore the original eigenvectors. Here we are particularly interested in the topological characteristics of the singularities on the two solid lines in Fig. 2b on which two or three EPs coalesce. For the convenience of discussion, we put the four states in equation (4) in the following order: $j=1:(-,+)$, $j=2:(-,-)$, $j=3:(+,-)$ and $j=4:(+,+)$, where the first (second) sign in the brackets denotes the choice of the first (second) sign in equation (4) outside (inside) the first square root and j=1 is the state with the lowest real frequency. In Fig. 5a, we plot the absolute value of phase rigidity[3], defined as $r_j = \langle \tilde{\phi}_j^R | \tilde{\phi}_j^R \rangle^{-1}$ for each state $j$ as a function of $\Delta\Gamma$ for a particular point on the red line of the phase diagram. Phase rigidity is a measure of the mixing of different states. In the absence of $\Delta\Gamma$, all four states are distinct and their phase rigidity is close to unity. As $\Delta\Gamma$ is increased, phase rigidities are reduced as some states start to mix. Two states are completely mixed at the EP where the phase rigidity vanishes. Clearly $|r_j|$ drops to zero at CS-1 ($\Delta\Gamma \cong 5.5$) for $j=2$ and 3. This indicates the existence of an EP at this point as state 2 and state 3 become linearly dependent with a vanishing norm $\langle \phi_{2,3}^L | \phi_{2,3}^R \rangle = 0$. In CS-3 ($\Delta\Gamma \cong 10.5$), all four states have zero rigidity, which indicates they are all linearly dependent with three defective states. In Fig. 5c, we also plot $|r_j|$ as a function of $|\Delta\Gamma - \Delta\Gamma_{CS-3}|$ using a dark-green line in log-log scale. A linear line with a slope of 3/4 is found. This corresponds to a higher order singularity resulting from

the coalescence of four states with three of them defective. To understand the exponent 3/4 physically, we have performed an adiabatic process encircling the CS-3 singularity in the complex $\Delta\Gamma$ plane in a counterclockwise direction as shown Fig. 5d. The imaginary part of $\Delta\Gamma$ represents a shift in the resonant frequency of the cavity. The trajectories of the four eigenfrequencies (real part) along the path are shown in Fig. 5e, from which we see that it requires four cycles to bring an eigenstate back to its original position. We have also calculated the geometric phase using parallel transport[29] and obtained a phase of $\pm 3\pi$ after four cycles. This indicates that eight cycles are required to restore the initial eigenvector. The log-log plot of $|r_j|$ near the CS-1 singularity at $\Delta\Gamma \cong 5.5$ gives an exponent of 1/2 as expected for an ordinary EP (blue line in Fig. 5c).

In Fig. 5b, we plot the phase rigidity for a particular point on the yellow line in the phase diagram. The figure shows that $|r_j|$ vanishes for states $j = 2$ and 3 at the linear crossing point at $\Delta\Gamma = 8$, indicating a defective state. However, the log-log plot of $|r_j|$ shown by the red curve in Fig. 5c gives an exponent of 1. This is different from the exponent of 1/2 for an ordinary EP and is also different from a diabolic point in a Hermitian Hamiltonian where no singularity is found. Thus, the yellow line represents a line consisting of EPs with a singularity different from that of an isolated EP. The trajectories of states 2 and 3 encircling the crossing point at $\Delta\Gamma = 8$ are shown in Fig. 5f, from which we find that only one cycle is required to bring the two states back to their original positions. The calculation of parallel transport gives a geometric phase of $\pm\pi$ after one cycle, consistent with the exponent of 1 found in Fig. 5c. To recap, we have observed for the first time higher order wavefunction singularities of exponents 1 and 3/4 due respectively to the coalescence of two EPs having the same chirality and the coalescence of three EPs.

**Conclusions**

The experimental system, as shown in Fig. 2a, can be viewed as a connected network of lossy cavities. The EP-related physics are expected to be even richer when the number of connected cavities is further increased where the symmetry of the network and the topology of the connectivity can serve as extra degrees of freedom. The new physics obtained here should also apply to electromagnetic and matter waves. As singularities underlie the essence of EPs, the new singularities found in higher dimensions and their associated topological properties can serve as new platforms for realizing new phenomena.

**Methods**

**Experiments.** The acoustic resonators were cylindrical cavities precision-machined out of stainless steel. All cavities had the same radius of 15.0 mm. To acquire the detuned eigenfrequencies for the fundamental mode, we fabricated in total eight cavities with four different heights: 50.0, 50.2, 50.4 and 50.6 mm. The eigenfrequency $\omega$ of the fundamental mode of the cavity can be tuned by varying its height $h$ through the relation $\omega = \pi v / h$, where $v$ is speed of sound in air. The eigenfrequency can be further fine-tuned by adding a small amount of Blu-Tack putty inside the cavity, which slightly decreases the volume. The cavities were filled with air at one atmospheric pressure, with temperature kept at 295 K. Small ports of 2.5 mm in diameter were opened on the top of the cavities for external pumping. This also introduced radiation loss that contributed to $\Gamma_0$. Side ports were fabricated to accommodate coupling tubes. A set of small tubes having the same length of 15.0 mm but various radii were also machined out of stainless steel. The couplings (both $\kappa$ and $t$) can be adjusted by choosing the tube's cross-sectional area. Microphones (PCB Piezotronics Model-378C10) were inserted into the wall of the cavities near the bottom. To introduce asymmetric loss, we stuck small pieces of sponge to the top/bottom of a cavity. However, this slightly red-shifted the resonant frequency. To counter this shift, we further added a small amount of Blu-Tack putty, which decreased the volume and consequently blue-shifted the resonant frequency. Therefore the mixture of sponge and putty yielded a total effect that ideally reproduced $\Delta\Gamma$, as shown in Fig. 1b. Lock-in amplifiers (Stanford Research SR-830) were used to drive the loudspeaker, as well as to record the signals from the microphones. For the four-cavity set-up, measurements were performed four times with the loudspeaker driving each cavity individually. The arithmetic mean of these four results yielded a spectrum under incoherent pumping.

The experimental realization of two connected acoustics cavities are shown in Fig. 1c, in which open dots show the measured pressure responses at cavity A (with pumping also at cavity A) compared with the fitted results obtained from a Green's function (see below). The inset of Fig. 1c shows the trajectories of two eigenfrequencies as $\Delta\Gamma$ is increased. We see that two eigen-modes coalesce at an EP when $\Delta\Gamma = 2|\kappa|$.

Experimental parameters for realizing the Class I EPFP as shown in Fig. 4(a) are as follows: The height of cavities A and B is 50.6 mm with 150 mg of putty inside. The height of cavities C and D is 50.0 mm. Two side tubes with lengths of 0.8 mm and 0.4 mm connect A to B and C to D to provide intra-pair coupling $\kappa$. Two tubes with the same radius of 1.2 mm connect A to D and B to C to provide inter-pair coupling $t$.

Experimental parameters for realizing the Class II EPFP as shown in Fig. 4(c) are as follows: The height of cavities A and B is 50.6 mm. The height of cavities C and D is 50.2 mm. Intra-pair coupling $\kappa$ is realized using two tubes 2.0 mm and 0.8 mm in radius. The two tubes providing $t$ both have a radius of 0.4 mm.

Experimental parameters for realizing the Class III EPFP as shown in Fig. 4(e) are as follows: The height of cavities A and B is 50.2 mm with 150 mg of putty inside. The height of cavities C and D is 50.0 mm. The two $\kappa$-tubes have radii of 2.0 mm and 0.8 mm. The two $t$-tubes have radii of 0.4 mm and 0.8 mm.

**Numerical fitting.** Using the eigenvalues and right/left eigenvectors of the Hamiltonian, we build up the Green's function of our system with $N$ states as

$$\ddot{G}(\omega) = \sum_{j=1}^{N} \frac{|\tilde{\phi}_j^R\rangle\langle\tilde{\phi}_j^L|}{\omega - \tilde{\omega}_j}, \tag{8}$$

where $|\tilde{\phi}_j^R\rangle$ and $\langle\tilde{\phi}_j^L|$ are the normalized biorthogonal right and left eigenvectors, and $\tilde{\omega}_j$ are eigenvalues. Then the analytical response function is $|P(\omega)| = A|\langle p|\ddot{G}(\omega)|s\rangle|$, where $|s\rangle$ and $|p\rangle$ are two column vectors describing the source and probe information. For example, in the two-cavity case (cavities A and B in Fig. 1a) the two basis vectors are $(1,0)^T$ for A and $(0,1)^T$ for B. By using experimental results measured without the putty-sponge mixture, the values of $\omega_{1,2}$, $\kappa$, $t$, and $\Gamma_0$ are obtained by fitting the measured $|P(\omega)|$. These parameters are kept fixed to fit $\Delta\Gamma$ by using the experimental results with the putty-sponge mixture added to induce loss.

**Acknowledgements**

G. M. thanks Ping Sheng for helpful discussions and Zhiyu Yang for providing equipment and lab space. This work was supported by the Hong Kong Research Grants Council (grant no. AoE/P-02/12).

**Author contributions**

K.D. performed the model calculations, and produced the theoretical results. G.M. designed and carried out the experiments. M.X. assisted in the theoretical and experimental analysis. Z.Q.Z. provided advice on the calculation and the analysis. C.T.C. led the research effort. All authors engaged in extensive discussions and co-wrote the paper.

**Competing Financial Interest statement**

The authors declare no competing financial interests.

**FIGURES AND FIGURE CAPTIONS**

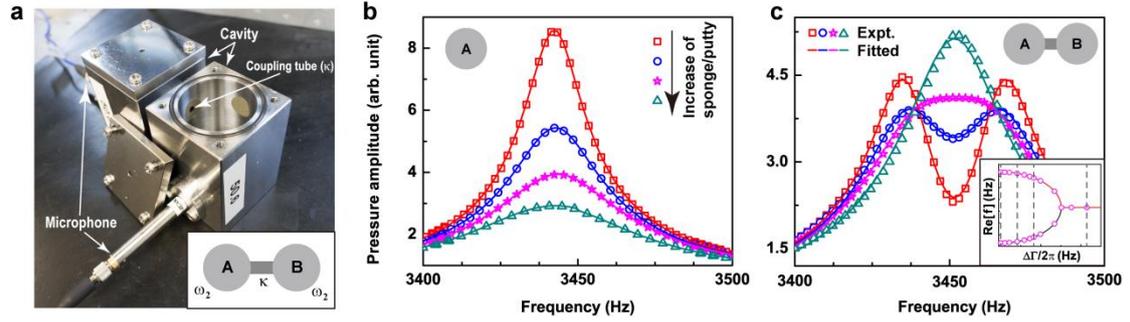

**Figure 1 | Experiments on single and double cavities. a,** A photograph of two acoustic cavity resonators (labeled A-B in the inset) coupled by side tubes. Cavity B is opened and disconnected to show its interior and the coupling tubes. The inset is a schematic picture of the system. **b,** Measured pressure responses as functions of frequency (markers) of a single cavity ( $h = 50.6$ mm ) with increasing loss. **c,** Measured pressure response spectra (markers) of the two coupled cavities. Cavities A and B have the same height $h = 50.6$ mm. Coupling $\kappa$ is achieved using two side tubes with radii of 2.0 mm and 0.8 mm, respectively. Asymmetric loss is introduced in cavity B, whereas both pumping and measurements are performed at cavity A. Solid curves in **b**, **c** are numerical fittings using a Green's function (see Methods). The inset in **c** shows fitted results of real parts of eigenfrequencies as a function of increasing loss. The vertical dash lines of the inset correspond to the four asymmetric loss $\Delta\Gamma$ shown in **c**. The formation of an exceptional point at which the two states coalesce can be clearly seen.

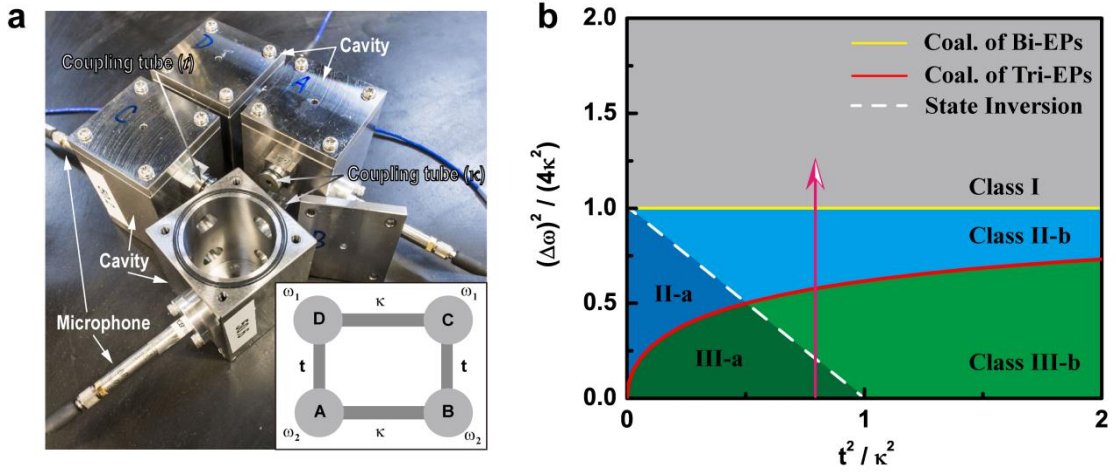

**Figure 2 | Phase diagram of a four-state system with asymmetric loss. a,** A photograph of four coupled acoustic cavity resonators (labeled A-D in the inset which shows a schematic drawing of the system). Here, A and B (C and D) form a pair with resonant frequency $\omega_2$ ($\omega_1$), with $\kappa$, $t$ being the coupling between these resonators. Phase diagram in the $\Delta\omega \sim t$ space is shown in **b**, with $\Delta\omega = \omega_1 - \omega_2$. In **b**, the grey, blue, and green regions represent Classes I, II, and III EPFPs, respectively. The solid red curve marks the coalescence of three EPs, the solid yellow line marks the coalescence of two EPs, and the white dashed line marks the state inversion line which separates subclasses "a" and "b" with different topological characteristics.

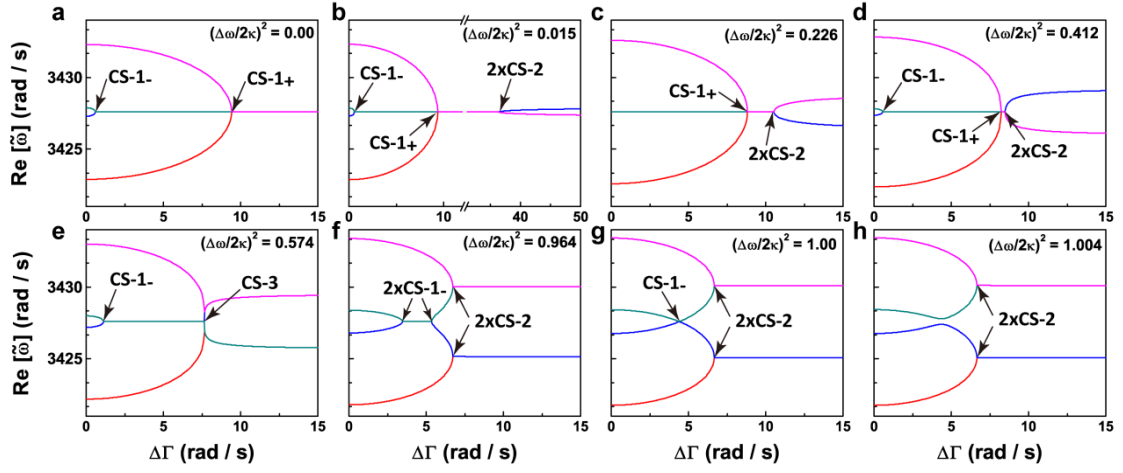

**Figure 3 | Evolution of eigenfrequency with increasing inter-pair frequency difference.** Real parts of eigenfrequencies with parameters $\omega_0 = 3427.59 \text{rad/s}$, $\kappa = -2.5 \text{rad/s}$, $t = -2.2 \text{rad/s}$, $\Gamma_0 = 10 \text{rad/s}$, and varying $\Delta\omega$ in the vertical red arrow shown in Fig. 2b are plotted in **a** to **h**. Starting from $(\Delta\omega/2\kappa)^2 = 0$ (**a**), gradually increasing $\Delta\omega$ will bring the system to Class III-a (**b**), the state inversion line (**c**), Class III-b (**d**), the red line of the CS-3 singularities (**e**), Class II-b (**f**), the yellow line of the coalescence of CS-1 singularities (**g**), and finally arriving at Class I (**h**).

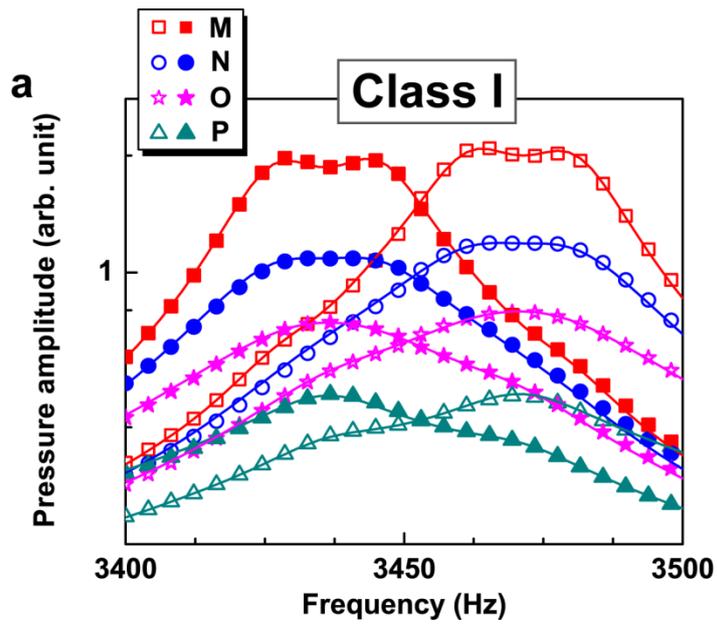
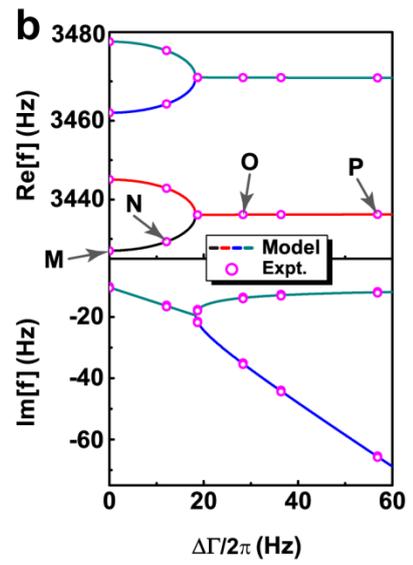
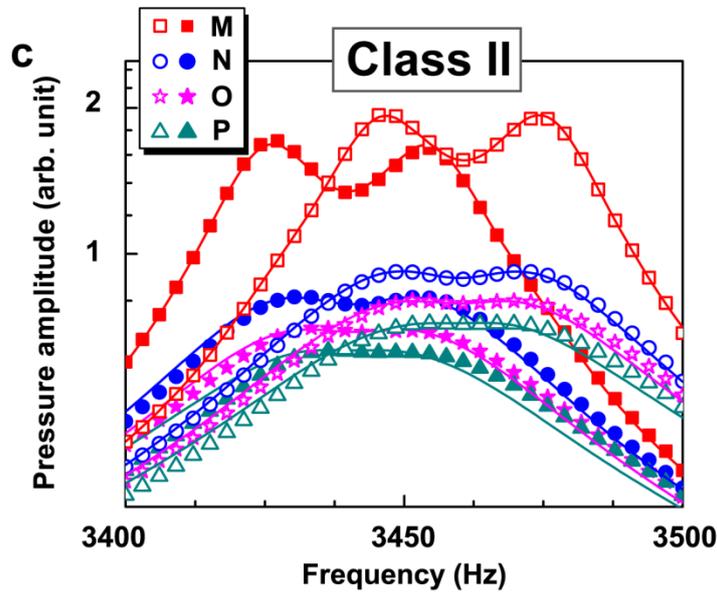
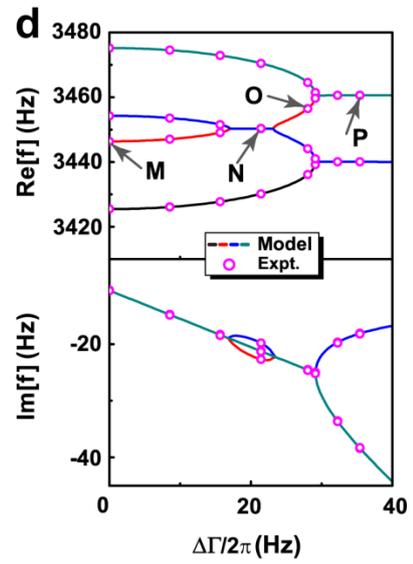
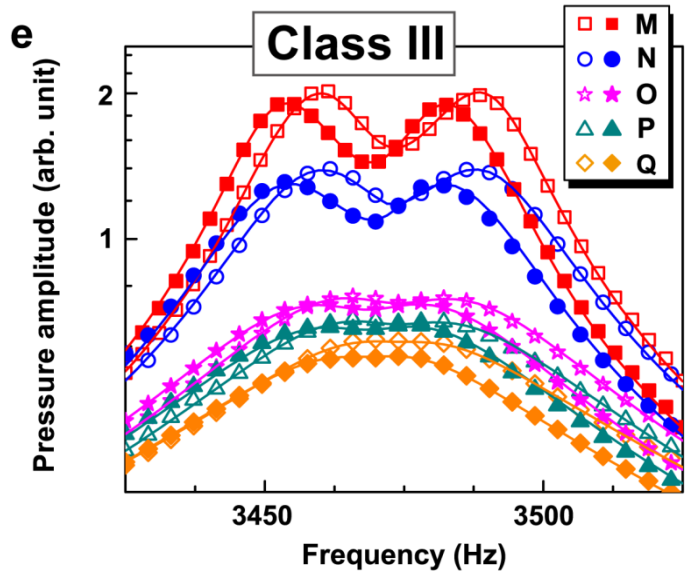
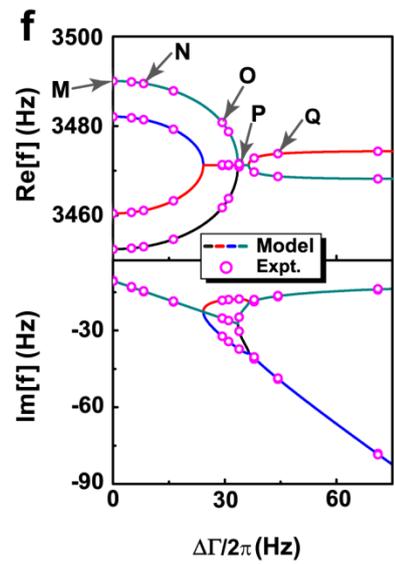

**Figure 4 | Experimental results of the three classes of exceptional point formation patterns for four cavities.** Measured pressure response spectra at cavity B (filled markers) and D (open makers) (**a, c, e**). The coupled cavities are pumped incoherently (see Methods). Results labeled M to P/Q shows increasing amounts of asymmetric loss. The corresponding solid curves are fitted using a Green's function (see Methods). (**b, d, f**) Real and imaginary parts of eigenfrequencies as functions of asymmetric loss $\Delta\Gamma/2\pi$ obtained by using parameters fitted from experimental results (markers) and analytical models (solid curves). The colors (black/red/blue/dark-cyan) represent different eigenfrequencies of our model.

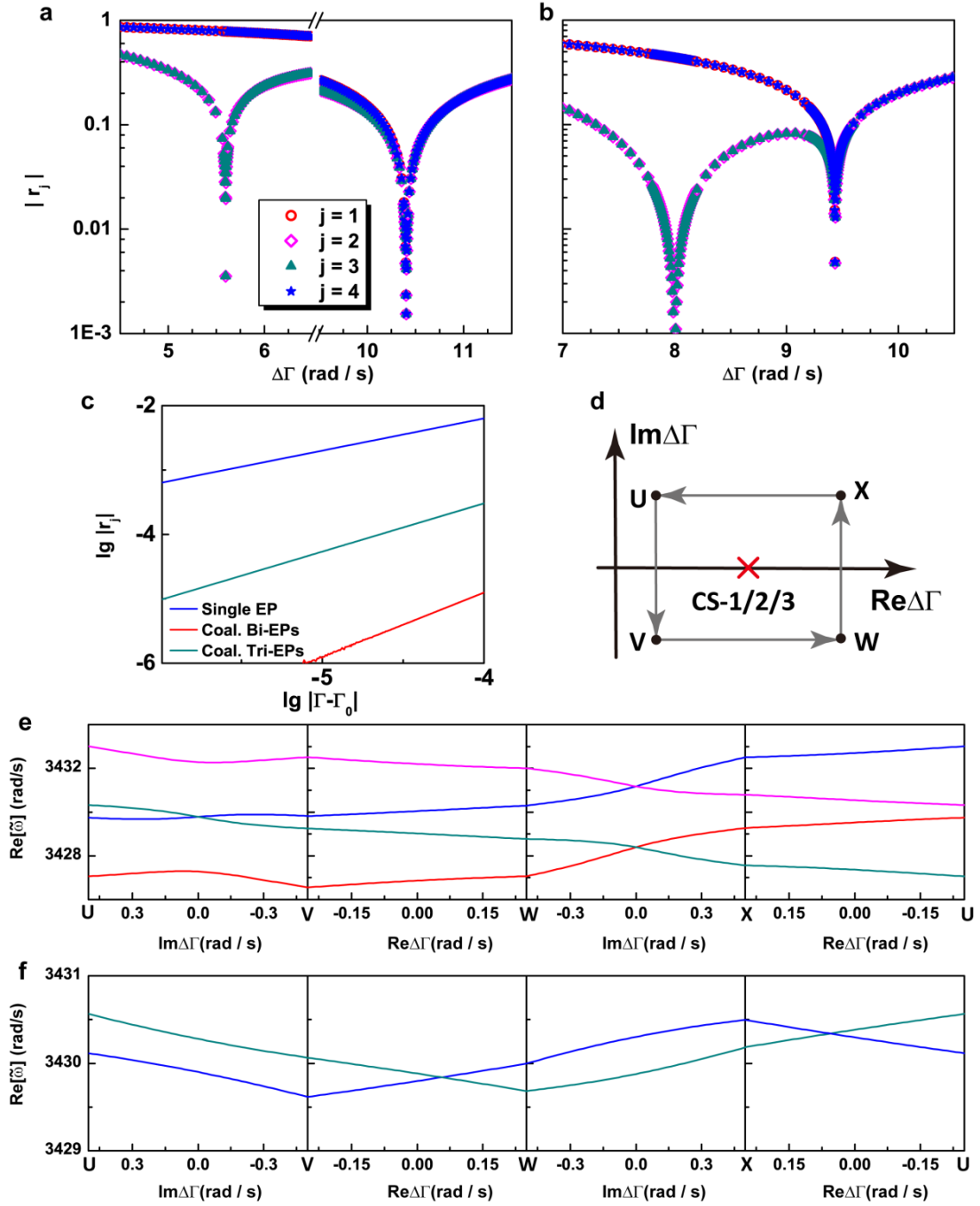

**Figure 5 | Phase rigidity and looping results of eigenstates. a,** Phase rigidity of all the eigenstates as functions of asymmetric loss $\Delta\Gamma$ for the parameters $\omega_1 = 3431.9746\text{rad/s}$, $\omega_2 = 3427.59\text{rad/s}$, $\kappa = -2.5\text{rad/s}$, $t = -4.0\text{rad/s}$, and $\Gamma_0 = 10\text{rad/s}$. **b,** Phase rigidity of all the eigenstates as functions of $\Delta\Gamma$ for the parameters $\omega_1 = 3432.59\text{rad/s}$, $\omega_2 = 3427.59\text{rad/s}$, $\kappa = -2.5\text{rad/s}$, $t = -4.0\text{rad/s}$, and $\Gamma_0 = 10\text{rad/s}$. **c,** Log-log plot of phase rigidity $|r_j|$ and $|\Delta\Gamma - \Delta\Gamma_0|$ for a single EP (blue curves), a CS-3 singularity (dark-cyan curve) and coalescence of two

CS-1 singularities (red curve). **d,** Looping path in the complex-$\Delta\Gamma$ plane, in which a CS-i singularity locates inside the loop. **e,** Eigenvalue trajectories for looping around a CS-3 singularity in the counterclockwise direction ($U \rightarrow V \rightarrow W \rightarrow X \rightarrow U$) shown in **d**. **f,** Eigenvalue trajectories for looping around the coalescence point of two CS-1 singularities in the counterclockwise direction ($U \rightarrow V \rightarrow W \rightarrow X \rightarrow U$) shown in **d**.

# Supplementary Information --- The emergence, coalescence and topological properties of multiple exceptional points and their experimental realization

Kun Ding[†], Guancong Ma[†], Meng Xiao, Z. Q. Zhang, and C. T. Chan[★]

*Department of Physics and Institute for Advanced Study,*
*The Hong Kong University of Science and Technology, Hong Kong*

[†]These authors contributed equally.
[★]Corresponding E-mail: phchan@ust.hk

## 1 Basic model

The Hamiltonian used to describe an acoustic system which composed of two pairs of acoustic cavities (for simplification, we assume $\omega_1 \geq \omega_2$) is

$$\mathrm{H} = \begin{pmatrix} \omega_2 - i\Gamma_0 & \kappa & & t \\ \kappa & \omega_2 - i\Gamma & t & \\ & t & \omega_1 - i\Gamma_0 & \kappa \\ t & & \kappa & \omega_1 - i\Gamma \end{pmatrix} = \mathrm{H}_0 + \mathrm{V}, \qquad (S1)$$

in which $\Delta\Gamma = \Gamma - \Gamma_0$,

$$\mathrm{H}_0 = \begin{pmatrix} \omega_2 - i\Gamma_0 & \kappa & & t \\ \kappa & \omega_2 - i\Gamma_0 & t & \\ & t & \omega_1 - i\Gamma_0 & \kappa \\ t & & \kappa & \omega_1 - i\Gamma_0 \end{pmatrix}, \quad \mathrm{V} = \begin{pmatrix} & & & \\ & -i\Delta\Gamma & & \\ & & & \\ & & & -i\Delta\Gamma \end{pmatrix}. \qquad (S2)$$

We note that $\kappa < 0, t < 0$ as the lower frequency mode is the even mode. To see the physics, we could rotate H to diagonal representation of $\mathrm{H}_0$,

$$\mathrm{H} = \left(\omega_0 - i\frac{\Gamma + \Gamma_0}{2}\right)\breve{\mathrm{I}} + \begin{pmatrix} -|\kappa| - \frac{1}{2}\sqrt{\Theta} & -i\frac{\Delta\Gamma}{2}\frac{\Delta\omega}{\sqrt{\Theta}} & & -i\Delta\Gamma\frac{|t|}{\sqrt{\Theta}} \\ -i\frac{\Delta\Gamma}{2}\frac{\Delta\omega}{\sqrt{\Theta}} & |\kappa| - \frac{1}{2}\sqrt{\Theta} & -i\Delta\Gamma\frac{|t|}{\sqrt{\Theta}} & \\ & -i\Delta\Gamma\frac{|t|}{\sqrt{\Theta}} & -|\kappa| + \frac{1}{2}\sqrt{\Theta} & i\frac{\Delta\Gamma}{2}\frac{\Delta\omega}{\sqrt{\Theta}} \\ -i\Delta\Gamma\frac{|t|}{\sqrt{\Theta}} & & i\frac{\Delta\Gamma}{2}\frac{\Delta\omega}{\sqrt{\Theta}} & |\kappa| + \frac{1}{2}\sqrt{\Theta} \end{pmatrix}, \qquad (S3)$$

in which $\omega_0 = (\omega_1 + \omega_2)/2$, $\Delta\omega = \omega_1 - \omega_2$, and $\Theta = 4t^2 + (\Delta\omega)^2$. Diagonalization of Eq.(S3) gives the eigen-frequencies as

$$\tilde{\omega}_j = \omega_0 - i\frac{\Gamma + \Gamma_0}{2} \pm \frac{1}{2}\sqrt{\Delta_1 \pm 4\sqrt{\Delta_2}} \qquad j = 1, 2, 3, 4, \qquad (S4)$$

in which the order of eigen-frequencies follow as (in increasing values of real frequency) $j=1:(-,+)$, $j=2:(-,-)$, $j=3:(+,-)$, $j=4:(+,+)$, and two kernel elements are

$$\Delta_1 = -(\Delta\Gamma)^2 + 4\kappa^2 + 4t^2 + (\Delta\omega)^2, \quad (S5)$$

$$\Delta_2 = 4\kappa^2 t^2 + \kappa^2(\Delta\omega)^2 - (\Delta\Gamma)^2 \frac{(\Delta\omega)^2}{4}. \quad (S6)$$

It is easy to see that what determine the properties of Eq.(S4) are intra-pair loss difference $\Delta\Gamma$, intra-pair coupling $\kappa$, inter-pair coupling $t$, and inter-pair frequency difference $\Delta\omega$. In additional to eigen-values, we also need to calculate eigenvectors of Hamiltonian (S1). There are two sets of eigen-vectors for non-herimitian matrix, namely right and left eigen-vectors

$$\begin{array}{ll} \text{Right:} & H\phi^R = \omega\phi^R \\ \text{Left:} & \phi^L H = \omega\phi^L \end{array}, \quad (S7)$$

where $\phi^R$ are column vector and $\phi^L$ are row vector. Since normalization condition for non-hermitian Hamiltonian is a bilinear product, so eigenvectors are normalized as

$$|\tilde{\phi}^R\rangle = \frac{|\phi^R\rangle}{\sqrt{\langle\phi^L|\phi^R\rangle}}, \quad \langle\tilde{\phi}^L| = \frac{\langle\phi^L|}{\sqrt{\langle\phi^L|\phi^R\rangle}}, \quad (S8)$$

and then right and left eigenvectors obey the following orthonormal condition

$$\langle\tilde{\phi}_i^L|\tilde{\phi}_j^R\rangle = \delta_{ij}. \quad (S9)$$

## 2 Bench mark: two decoupled pairs

If we set $t=0$, then the two pairs of cavities become decoupled, which means Hamiltonian (S1) is block diagonalized, so eigen-values of Hamiltonian could be easily given out

$$\tilde{\omega}_j = \omega_{2,1} - i\frac{\Gamma+\Gamma_0}{2} \pm \frac{1}{2}\sqrt{4\kappa^2 - (\Delta\Gamma)^2}. \quad (S10)$$

It is easy to see the square-root singularity in the spectrum. To show this singularity, we plot eigen-frequencies for only one pair in (S10), as shown in Figs. S1(a), S1(c) and S1(d).

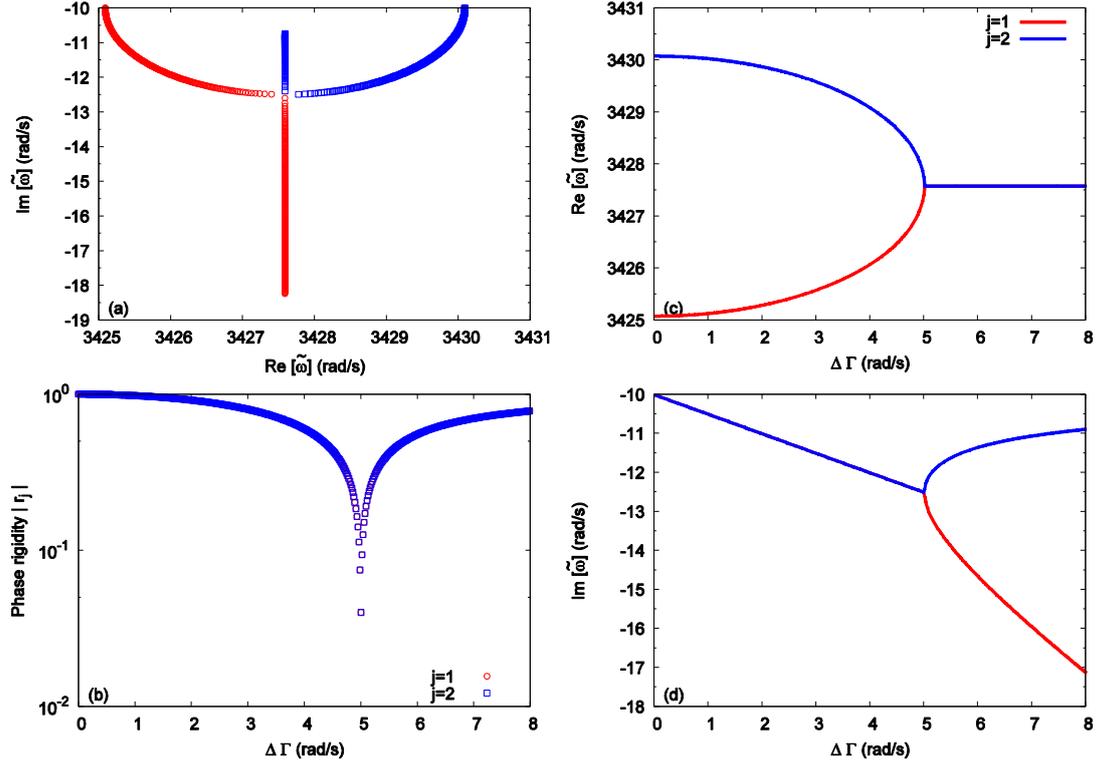

**Figure S1.** (a) Trajectory of eigen-frequency $\tilde{\omega}_j$ in the complex frequency plane. (b) Norm of phase rigidity $|r_j|$ for each state as function of loss difference $\Delta\Gamma$. (c) Real parts and (d) imaginary parts of eigen-frequency $\tilde{\omega}_j$ as function of loss difference $\Delta\Gamma$. Parameters used here are $\omega_1 = 3427.59 \text{Hz}$, $\kappa = -2.5\text{Hz}$, and $\Gamma_0 = 10\text{Hz}$.

One way to characterize the phase transition at the EP is to study the phase rigidity of the two states, which is defined as[1]

$$r_j = \frac{\langle \tilde{\phi}_j^L | \tilde{\phi}_j^R \rangle}{\langle \tilde{\phi}_j^R | \tilde{\phi}_j^R \rangle} = \frac{1}{\langle \tilde{\phi}_j^R | \tilde{\phi}_j^R \rangle} . \tag{S11}$$

The results are shown in Fig. S1(b). It is clearly seen that $0 \leq |r_j| \leq 1$, and phase rigidity approaches zero. The completely overlap of the two curves at EP is a manifestation of two linearly dependent states at EP, i.e, one state is defective.

Another way to characterize an EP is to trace the evolution of eigenstates under some parameter variation, i.e., $|\tilde{\phi}_j^R(\Delta\Gamma)\rangle$ [1]. By projecting each eigenfunction ($j = 1, 2$) onto the two original states when $\Delta\Gamma = 0$, i.e., $a_{jk} = \langle \tilde{\phi}_{(0),k}^L | \tilde{\phi}_j^R(\Delta\Gamma) \rangle$, where $k = 1, 2$. The subscript "(0)" stands for the eigenstates when $\Delta\Gamma = 0$. The results are shown in Fig. S2, from which we observe clearly the mixing of two states as EP is approached and complete mixing after EP.

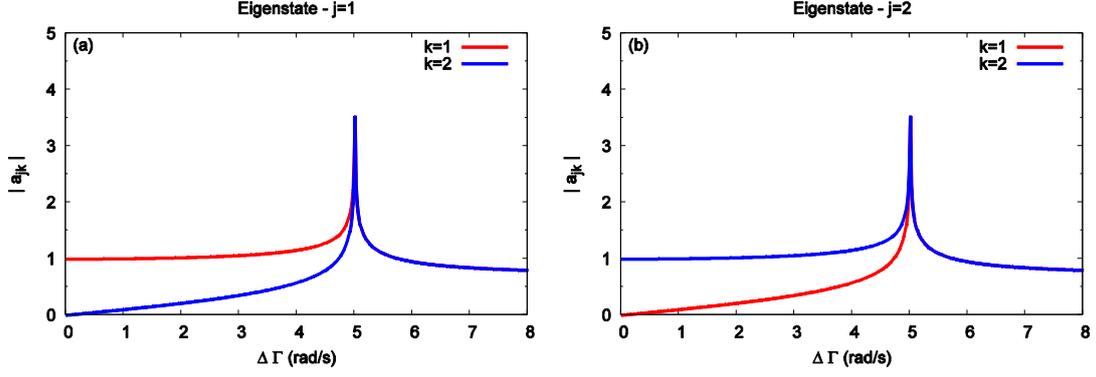

**Figure S2.** Evolution of eigenstate-$j$ in Figure S1 as function of loss difference $\Delta\Gamma$ in the $\Gamma_0$-representation, in which $a_{jk} = \langle \phi_{(0),k}^L | \phi_j^R(\Delta\Gamma) \rangle$, and $\{\langle \phi_{(0),k}^L |; k=1,2\}$ are eigenstates at $\Delta\Gamma = 0$.

The complete mixing of two states after EP certainly does not mean that two states become identical. Their difference can be seen if we project the eigenfunction onto the cavity representation. The results are shown in Fig. S3. It's seen that after EP, the state with a larger imaginary frequency ($j=1$) is more concentrated in Cavity B and the other state with a smaller imaginary frequency ($j=2$) is more concentrated in Cavity A because the loss is more in Cavity B.

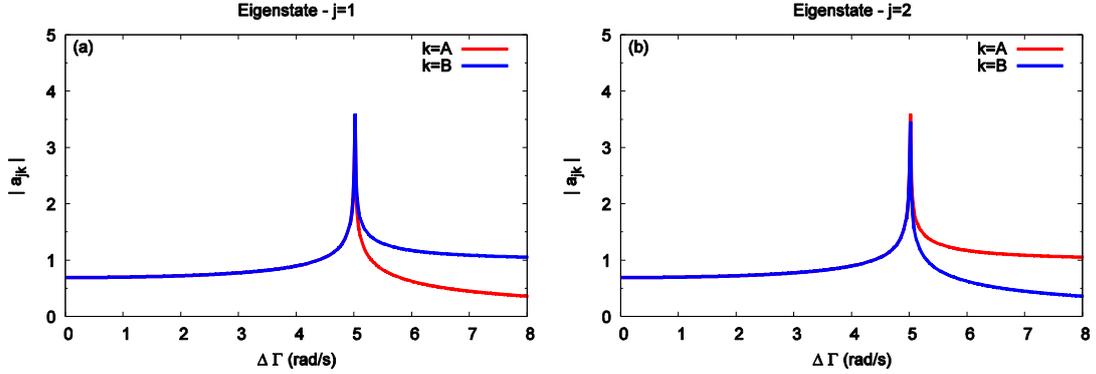

**Figure S3.** Evolution of eigenstate-$j$ in Figure S1 as function of loss difference $\Delta\Gamma$ in the cavity-representation, in which $a_{jk} = \langle \phi_{(0),k}^L | \phi_j^R(\Delta\Gamma) \rangle$, and $\{\langle \phi_{(0),k}^L |; k=\text{A,B}\}$ are unit basis vectors.

## 3 The three special lines in the parameter space

For the special case with $\Delta\Gamma = 0$, states 2 and state 3 could have state inversions (switching of the ordering of the frequency), and $H_{22} = H_{33}$ defines this state inversion line as

$$4\kappa^2 = 4t^2 + (\Delta\omega)^2. \tag{S12}$$

As mentioned in the main text, Eq.(S4) gives four cases when coalescence of states (CS) can occur, i.e., (CS-1): $\Delta_1 \pm 4\sqrt{\Delta_2} = 0$, (CS-2): $\Delta_2 = 0, \Delta_1 \neq 0$ and (CS-3): $\Delta_1 = \Delta_2 = 0$. If we denote $F_\pm \equiv \Delta_1 \pm 4\sqrt{\Delta_2}$, the mathematical conditions under which the coalescence of two CS-1 singularities could happen are $F_- = 0$ and $\frac{\partial F_-}{\partial \Delta\Gamma} = 0$, namely

$$\begin{cases} F_- = \Delta_1 - 4\sqrt{\Delta_2} = 0 \\ \frac{\partial F_-}{\partial \Gamma} = -2\Delta\Gamma\left[1 - \frac{(\Delta\omega)^2}{2\sqrt{\Delta_2}}\right] = 0 \end{cases}. \quad (S13)$$

Solving Eq.(S13) gives the following condition

$$t^2\left[(\Delta\omega)^2 - 4\kappa^2\right] = 0 \Rightarrow t = 0 \quad \text{or} \quad (\Delta\omega)^2 = 4\kappa^2. \quad (S14)$$

Solving CS-3 singularity condition $\Delta_1 = \Delta_2 = 0$ gives the following equation (we can call them "kissing lines" from a phenomenal point of view)

$$(\Delta\omega)^2_{ks} = 2\sqrt{t^4 + 4\kappa^2 t^2} - 2t^2, \quad (S15)$$

$$\Delta\Gamma^2_{ks} = 2|t|\sqrt{t^2 + 4\kappa^2} + 4\kappa^2 + 2t^2. \quad (S16)$$

If $|t| \to 0$, then $(\Delta\omega)_{ks} = \sqrt{4|\kappa t|}$. And if $|t| \to \infty$, then $(\Delta\omega)^2_{ks} = 4\kappa^2$. Eqs.(S12) (S14) and (S15) give out three special lines in the parameter space, as shown in Fig.2b of main text. In the following sections, we will discuss the properties of the system in different limits and discuss the topological properties of these lines.

## 4 Case one: $(2|t|/\Delta\omega) \ll 1$

Physically, this case means the inter-pair coupling is much smaller than inter-pair frequency difference, so we could expand $t$ in Eq.(S3) to leading order,

$$H \approx \begin{pmatrix} \omega_2 - |\kappa| - i\frac{\Gamma + \Gamma_0}{2} & -i\frac{\Delta\Gamma}{2} & & \\ -i\frac{\Delta\Gamma}{2} & \omega_2 + |\kappa| - i\frac{\Gamma + \Gamma_0}{2} & & \\ & & \omega_1 - |\kappa| - i\frac{\Gamma + \Gamma_0}{2} & i\frac{\Delta\Gamma}{2} \\ & & i\frac{\Delta\Gamma}{2} & \omega_1 + |\kappa| - i\frac{\Gamma + \Gamma_0}{2} \end{pmatrix}.$$ (S17)

$$+ \left(\frac{|t|}{\Delta\omega}\right) \ddot{\mathrm{I}} \begin{pmatrix} & & & -i\Delta\Gamma \\ & & -i\Delta\Gamma & \\ & -i\Delta\Gamma & & \\ -i\Delta\Gamma & & & \end{pmatrix}$$

It is not difficult to see from Eq.(S17) that if $t = 0$, two pairs become decoupled, so we could write down eigen-frequencies from Eq.(S4)

$$\tilde{\omega}_j = \omega_0 - i\frac{\Gamma + \Gamma_0}{2} \pm \frac{1}{2}\left(\Delta\omega \pm \sqrt{4\kappa^2 - \Delta\Gamma^2}\right).$$ (S18)

Furthermore, we could write Eq.(S18) explicitly

$$\tilde{\omega}_1 = \omega_2 - i\frac{\Gamma + \Gamma_0}{2} - \frac{1}{2}\sqrt{4\kappa^2 - \Delta\Gamma^2},$$ (S19)

$$\tilde{\omega}_2 = \omega_2 - i\frac{\Gamma + \Gamma_0}{2} + \frac{1}{2}\sqrt{4\kappa^2 - \Delta\Gamma^2},$$ (S20)

$$\tilde{\omega}_3 = \omega_1 - i\frac{\Gamma + \Gamma_0}{2} - \frac{1}{2}\sqrt{4\kappa^2 - \Delta\Gamma^2},$$ (S21)

$$\tilde{\omega}_4 = \omega_1 - i\frac{\Gamma + \Gamma_0}{2} + \frac{1}{2}\sqrt{4\kappa^2 - \Delta\Gamma^2}.$$ (S22)

Eigen states 2 and 3 could cross over each other at particular $\Delta\Gamma$ as

$$\Delta\Gamma \equiv \Gamma_d = \sqrt{4\kappa^2 - (\Delta\omega)^2}.$$ (S23)

Such crossover exists when $(\Delta\omega)^2 < 4\kappa^2$ which means intra-pair coupling is strong enough to create a mode inversion. Under such conditions, we could expand $\Delta\Gamma = \Gamma_d + \delta\Gamma$ and $t$ spontaneously in Eq.(S17) to their leading order, and then the reduced $2 \times 2$ Hamilton for state 2 and 3 becomes,

$$\begin{aligned} H_{red} &= \left[\omega_0 - \frac{i}{2}(2\Gamma_0 + \Gamma_d + \delta\Gamma)\right]\ddot{\mathrm{I}} + \left(\frac{\Gamma_d}{2\Delta\omega}\right)\ddot{\mathrm{I}} \begin{pmatrix} \delta\Gamma & i2|t|\mathrm{sgn}(\Delta\omega - 2\kappa) \\ i2|t|\mathrm{sgn}(\Delta\omega - 2\kappa) & -\delta\Gamma \end{pmatrix} \\ &= \left[\omega_0 - \frac{i}{2}(2\Gamma_0 + \Gamma_d + \delta\Gamma)\right]\ddot{\mathrm{I}} + \left(\frac{\Gamma_d}{2\Delta\omega}\right)\ddot{\mathrm{I}}\left(\delta\Gamma\sigma_z + i2|t|\sigma_x\right) \end{aligned}.$$ (S24)

The kernel in the bracket determines the bifurcation. We can rotate the kernel to our familiar representation, i.e.

$$H_{eff} = -\delta\Gamma\sigma_x + i2|t|\sigma_z. \tag{S25}$$

It is easy to see that inter-pair coupling plays the role of an "imaginary mass" term. Furthermore, eigen-values are

$$\tilde{\omega}_{2,3} = \omega_0 - \frac{i}{2}(2\Gamma_0 + \Gamma_d + \delta\Gamma) \pm \left(\frac{\Gamma_d}{2\Delta\omega}\right)\sqrt{(\delta\Gamma)^2 - 4t^2}, \tag{S26}$$

eigen-vectors are

$$\vec{a}_{\pm} = \begin{pmatrix} i\left(-\dfrac{\delta\Gamma}{2|t|} \mp \sqrt{(\dfrac{\delta\Gamma}{2t})^2 - 1}\right) \\ 1 \end{pmatrix}. \tag{S27}$$

So the chirality of two EPs is right-handed and left-handed respectively. To illustrate the properties of this case, we plot eigen-frequencies, phase rigidity and evolution of eigenstates in Figs. S4, and S5.

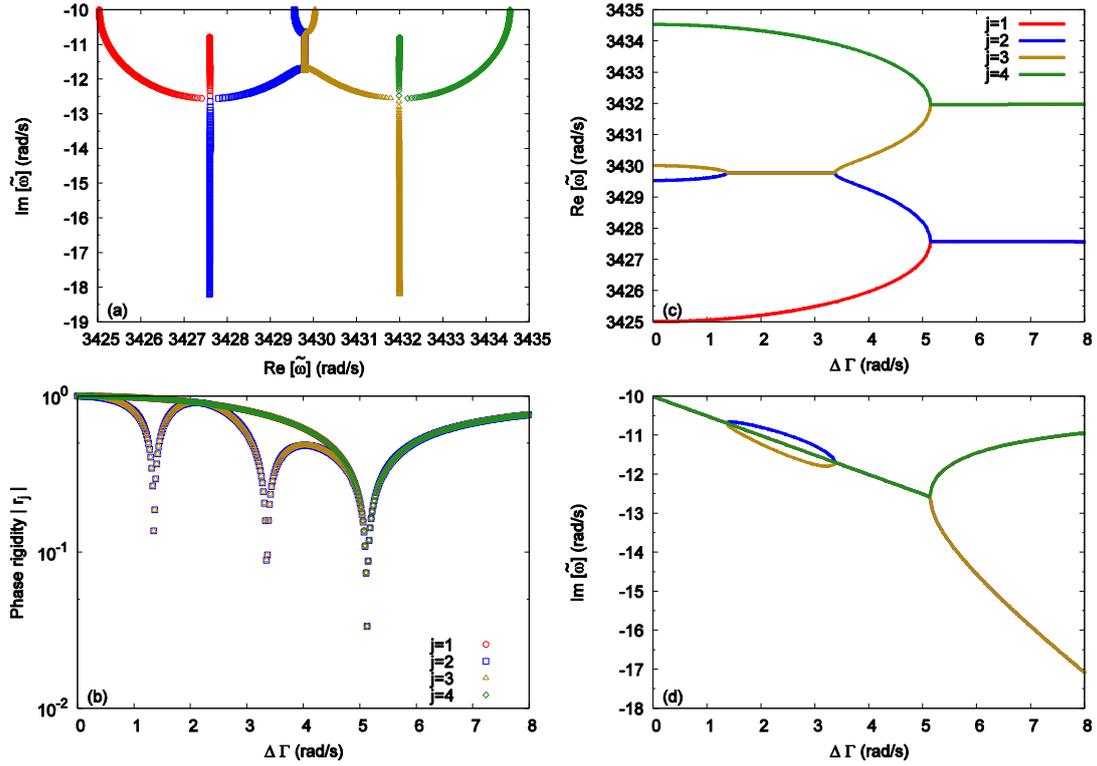

**Figure S4.** (a) Trajectory of eigen-frequency $\tilde{\omega}_j$ in the complex frequency plane. (b) Norm of phase rigidity $|r_j|$ for each state as function of loss difference $\Delta\Gamma$. (c) Real parts and (d) imaginary parts of eigen-frequency $\tilde{\omega}_j$ as function of loss difference $\Delta\Gamma$. Parameters used here are $\omega_1 = 3427.59\text{Hz}$, $\omega_2 = 3432.0\text{Hz}$, $\kappa = -2.5\text{Hz}$, $t = -0.5\text{Hz}$, and $\Gamma_0 = 10\text{Hz}$.

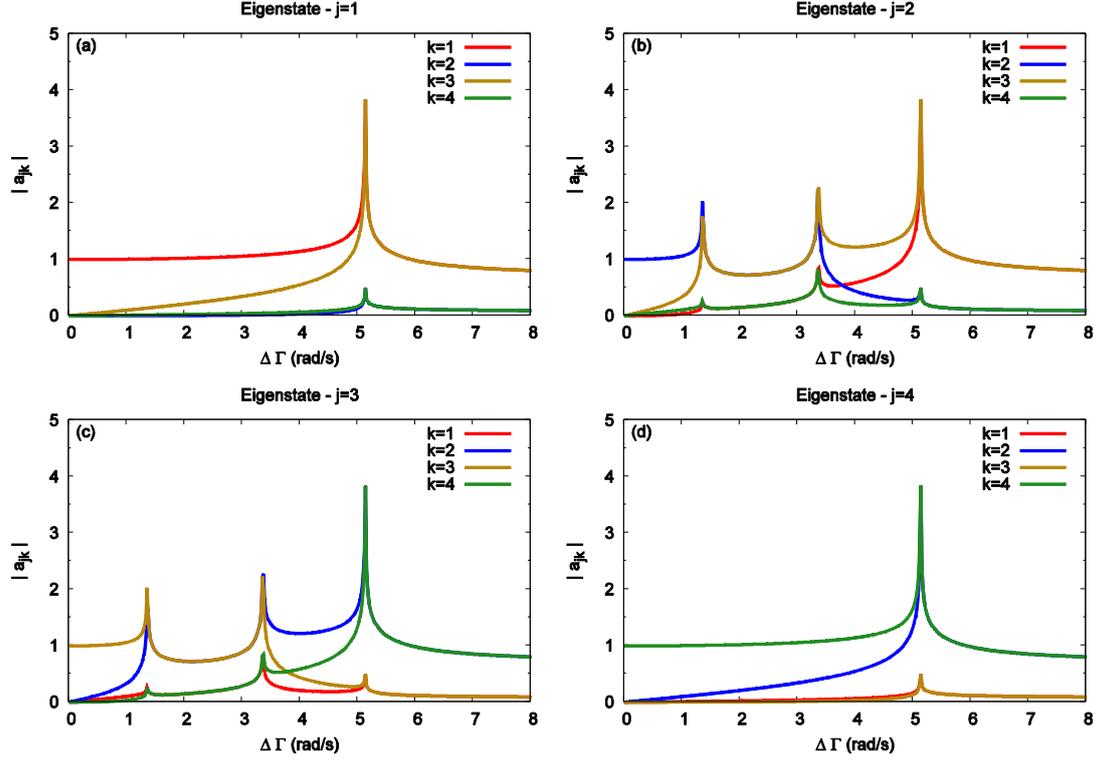

**Figure S5.** Evolution of eigenstate-$j$ in Figure S4 as function of loss difference $\Delta\Gamma$ in the $\Gamma_0$-representation, in which $a_{jk} = \langle \phi^L_{(0),k} | \phi^R_j(\Delta\Gamma) \rangle$, and $\{\langle \phi^L_{(0),k} |; k=1,2,3,4\}$ are eigenstates at $\Delta\Gamma = 0$.

## 5 Case two: $(\Delta\omega)^2 \approx 4\kappa^2$

In this case, we expand our system as

$$\frac{(\Delta\omega)^2}{4\kappa^2} = 1+\alpha, \qquad |\alpha| \ll 1. \tag{S28}$$

Then we could expand $\alpha$ in Eq.(S3) to leading order,

$$H \approx \begin{pmatrix} \omega_0 - |\kappa| - i\frac{\Gamma+\Gamma_0}{2} - \sqrt{t^2+\kappa^2} & -i\frac{\Delta\Gamma}{2}\frac{|\kappa|}{\sqrt{t^2+\kappa^2}} & & -i\frac{\Delta\Gamma}{2}\frac{|t|}{\sqrt{t^2+\kappa^2}} \\ -i\frac{\Delta\Gamma}{2}\frac{|\kappa|}{\sqrt{t^2+\kappa^2}} & \omega_0 + |\kappa| - i\frac{\Gamma+\Gamma_0}{2} - \sqrt{t^2+\kappa^2} & -i\frac{\Delta\Gamma}{2}\frac{|t|}{\sqrt{t^2+\kappa^2}} & \\ & -i\frac{\Delta\Gamma}{2}\frac{|t|}{\sqrt{t^2+\kappa^2}} & \omega_0 - |\kappa| - i\frac{\Gamma+\Gamma_0}{2} + \sqrt{t^2+\kappa^2} & i\frac{\Delta\Gamma}{2}\frac{|\kappa|}{\sqrt{t^2+\kappa^2}} \\ -i\frac{\Delta\Gamma}{2}\frac{|t|}{\sqrt{t^2+\kappa^2}} & & i\frac{\Delta\Gamma}{2}\frac{|\kappa|}{\sqrt{t^2+\kappa^2}} & \omega_0 + |\kappa| - i\frac{\Gamma+\Gamma_0}{2} + \sqrt{t^2+\kappa^2} \end{pmatrix}$$

$$+ \begin{pmatrix} -\frac{\kappa^2}{2\sqrt{t^2+\kappa^2}}\alpha & -i\frac{\Delta\Gamma}{4}\frac{|\kappa|t^2}{(\kappa^2+t^2)^{3/2}}\alpha & & i\frac{\Delta\Gamma}{4}\frac{|t|\kappa^2}{(\kappa^2+t^2)^{3/2}}\alpha \\ -i\frac{\Delta\Gamma}{4}\frac{|\kappa|t^2}{(\kappa^2+t^2)^{3/2}}\alpha & -\frac{\kappa^2}{2\sqrt{t^2+\kappa^2}}\alpha & i\frac{\Delta\Gamma}{4}\frac{|t|\kappa^2}{(\kappa^2+t^2)^{3/2}}\alpha & \\ & i\frac{\Delta\Gamma}{4}\frac{|t|\kappa^2}{(\kappa^2+t^2)^{3/2}}\alpha & \frac{\kappa^2}{2\sqrt{t^2+\kappa^2}}\alpha & i\frac{\Delta\Gamma}{4}\frac{|\kappa|t^2}{(\kappa^2+t^2)^{3/2}}\alpha \\ i\frac{\Delta\Gamma}{4}\frac{|t|\kappa^2}{(\kappa^2+t^2)^{3/2}}\alpha & & i\frac{\Delta\Gamma}{4}\frac{|\kappa|t^2}{(\kappa^2+t^2)^{3/2}}\alpha & \frac{\kappa^2}{2\sqrt{t^2+\kappa^2}}\alpha \end{pmatrix}$$

(S29)

When $\alpha = 0$, the system does NOT become two decoupled pairs which is different from previous two cases, but eigen-frequencies still canbe written as

$$\tilde{\omega}_j = \omega_0 - i\frac{\Gamma+\Gamma_0}{2} \pm \frac{1}{2}\left|2|\kappa| \pm \sqrt{4t^2 + 4\kappa^2 - \Delta\Gamma^2}\right|. \tag{S30}$$

It is not difficult to see that when $(\Delta\Gamma)^2 = 4t^2$, $\tilde{\omega}_2 = \tilde{\omega}_3$. Denote $\Gamma_d^2 = 4t^2$, and do expansion to $\Delta\Gamma = \Gamma_d + \delta\Gamma$ while keeping $\alpha = 0$ gives out

$$\tilde{\omega}_{2,3} = \omega_0 - \frac{i}{2}(2\Gamma_0 + \Gamma_d + \delta\Gamma) \pm \frac{|t|}{2|\kappa|}\delta\Gamma, \tag{S31}$$

which indicates there indeed a linear cross at $\Gamma_d$. Then do expansion to $\alpha$ while keeping $\delta\Gamma = 0$

$$\tilde{\omega}_{2,3} = \omega_0 - i(\Gamma_0 + |t|) \pm |t|\sqrt{\alpha}, \tag{S32}$$

which indicates that $\sqrt{\alpha}$ plays the role of mass. Finally we could do expansion to $\alpha$ and $\delta\Gamma$ spontaneously

$$\tilde{\omega}_{2,3} = \omega_0 - \frac{i}{2}(2\Gamma_0 + \Gamma_d + \delta\Gamma) \pm \frac{1}{2}\sqrt{\frac{t^2}{\kappa^2}(\delta\Gamma)^2 + 4t^2\alpha + 2|t|(1+\frac{t^2}{\kappa^2})(\delta\Gamma)\alpha}, \tag{S33}$$

which indicates that $\alpha > 0$ means "mass" is real while $\alpha < 0$ means "mass" is imaginary. To clarify this point, we plot eigen-states and phase rigidity of $\alpha = 0$ case (as shown in Figs. S6, and S7) and $\alpha > 0$ case (as shown in Figs. S8, and S9).

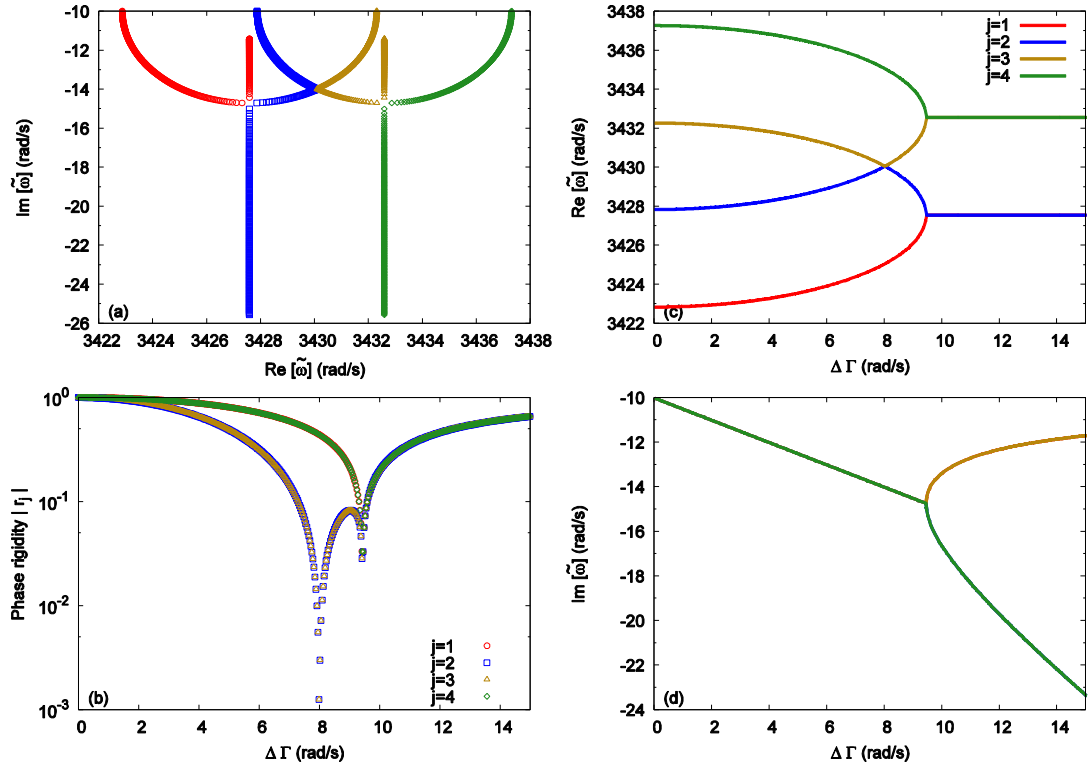

**Figure S6.** (a) Trajectory of eigen-frequency $\tilde{\omega}_j$ in the complex frequency plane. (b) Norm of phase rigidity $|r_j|$ for each state as function of loss difference $\Delta\Gamma$. (c) Real parts and (d) imaginary parts of eigen-frequency $\tilde{\omega}_j$ as function of loss difference $\Delta\Gamma$. Parameters used here are $\omega_1 = 3427.59\text{Hz}$, $\omega_2 = 3432.59\text{Hz}$, $\kappa = -2.5\text{Hz}$, $t = -4.0\text{Hz}$, and $\Gamma_0 = 10\text{Hz}$.

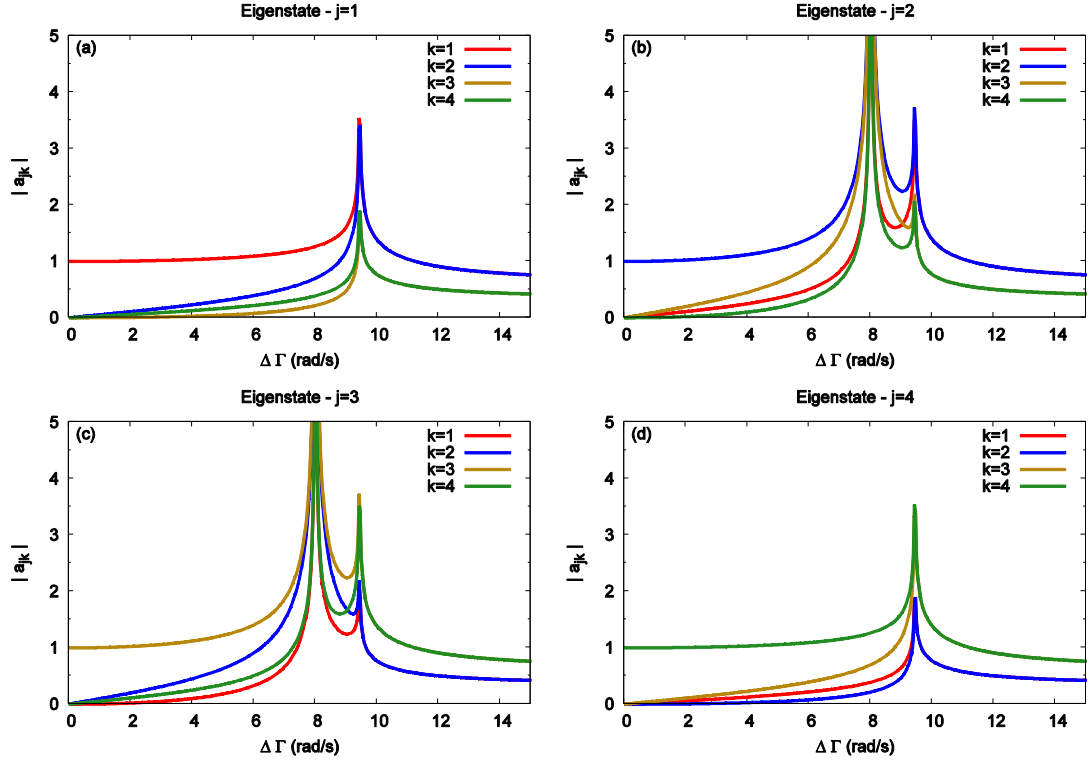

**Figure S7.** Evolution of eigenstate-$j$ in Figure S6 as function of loss difference $\Delta\Gamma$ in the $\Gamma_0$-representation, in which $a_{jk} = \langle \phi^L_{(0),k} | \phi^R_j(\Delta\Gamma) \rangle$, and $\{\langle \phi^L_{(0),k} |; k=1,2,3,4\}$ are eigenstates at $\Delta\Gamma = 0$.

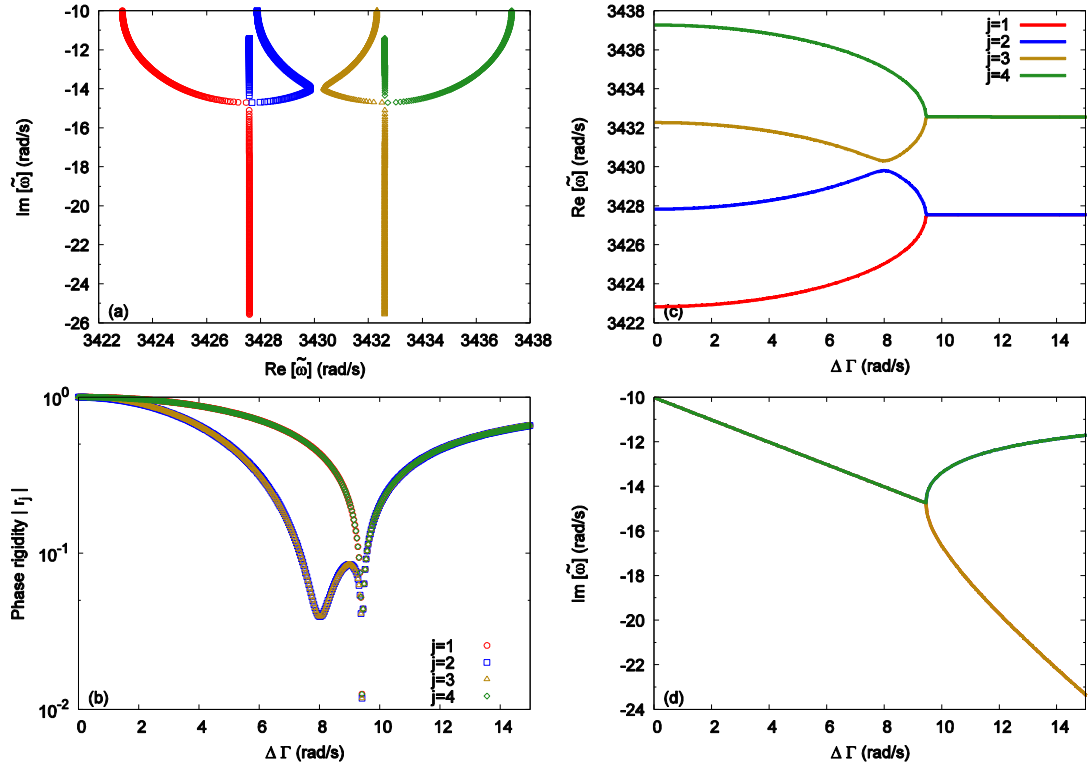

**Figure S8.** (a) Trajectory of eigen-frequency $\tilde{\omega}_j$ in the complex frequency plane. (b) Norm of phase rigidity $|r_j|$ for each state as function of loss difference $\Delta\Gamma$. (c) Real parts and (d) imaginary parts of eigen-frequency $\tilde{\omega}_j$ as function of loss difference $\Delta\Gamma$. Parameters used here are $\omega_1 = 3427.59\text{Hz}$, $\omega_2 = 3432.60\text{Hz}$, $\kappa = -2.5\text{Hz}$, $t = -4.0\text{Hz}$, and $\Gamma_0 = 10\text{Hz}$.

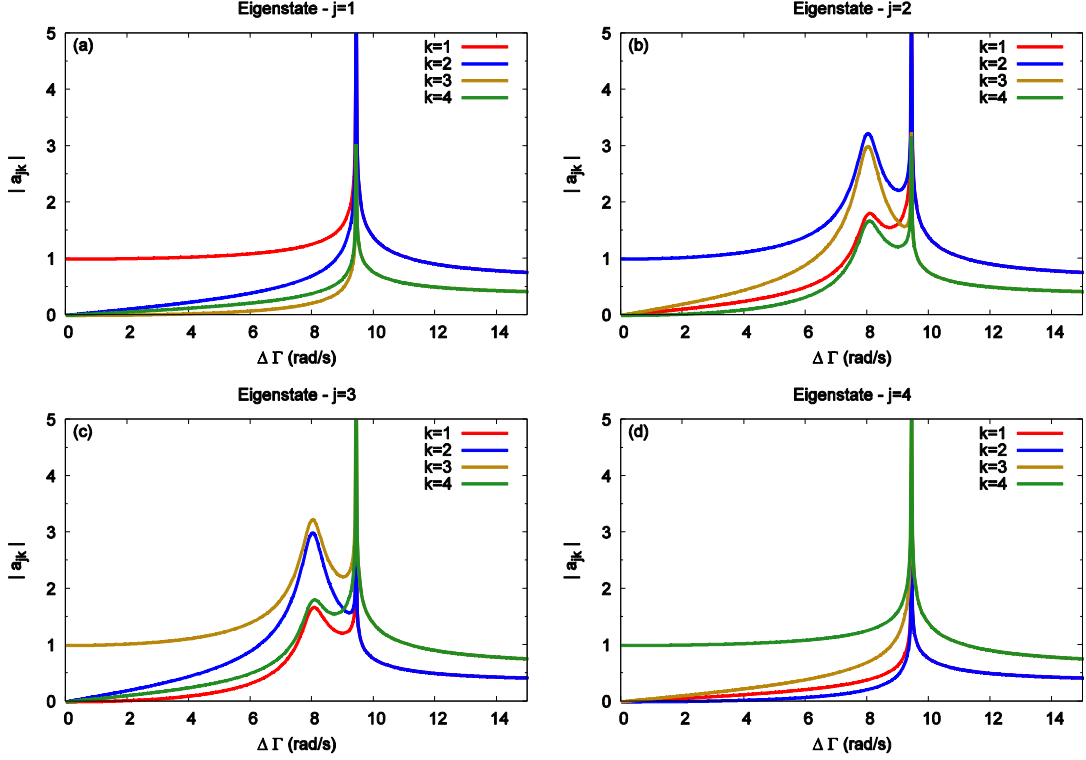

**Figure S9.** Evolution of eigenstate-$j$ in Figure S8 as function of loss difference $\Delta\Gamma$ in the $\Gamma_0$-representation, in which $a_{jk} = \langle \phi^L_{(0),k} | \phi^R_j(\Delta\Gamma) \rangle$, and $\{\langle \phi^L_{(0),k}|; k=1,2,3,4\}$ are eigenstates at $\Delta\Gamma = 0$.

## 6 Case three: $(\Delta\omega/2|t|) \ll 1$

Physically, this case means the inter-pair coupling is much larger than inter-pair frequency difference, so we could expand $\Delta\omega$ in Eq.(S3) to leading order,

$$H \approx \begin{pmatrix} \omega_0 - |\kappa| - i\frac{\Gamma+\Gamma_0}{2} - |t| & & & -i\frac{\Delta\Gamma}{2} \\ & \omega_0 + |\kappa| - i\frac{\Gamma+\Gamma_0}{2} - |t| & -i\frac{\Delta\Gamma}{2} & \\ & -i\frac{\Delta\Gamma}{2} & \omega_0 - |\kappa| - i\frac{\Gamma+\Gamma_0}{2} + |t| & \\ -i\frac{\Delta\Gamma}{2} & & & \omega_0 + |\kappa| - i\frac{\Gamma+\Gamma_0}{2} + |t| \end{pmatrix}$$

$$+ \begin{pmatrix} & & -i\frac{\Delta\Gamma}{2}\frac{\Delta\omega}{2|t|} & \\ -i\frac{\Delta\Gamma}{2}\frac{\Delta\omega}{2|t|} & & & \\ & & & i\frac{\Delta\Gamma}{2}\frac{\Delta\omega}{2|t|} \\ & i\frac{\Delta\Gamma}{2}\frac{\Delta\omega}{2|t|} & & \end{pmatrix}.$$

(S34)

When $\Delta\omega = 0$, the system also becomes two decoupled pairs, but EPs are different from case one. We could write down eigen-frequencies of Eq.(S34) from Eq.(S4),

$$\tilde{\omega}_j = \omega_0 - i\frac{\Gamma+\Gamma_0}{2} \pm \frac{1}{2}\sqrt{4(\kappa \pm t)^2 - \Delta\Gamma^2}.$$ (S35)

When $\Delta\omega \neq 0$, another two EPs come out at $\Delta_2 = 0$,

$$\Delta\Gamma_{CS-2}^2 = 4\kappa^2 \left(1 + \frac{4t^2}{(\Delta\omega)^2}\right).$$ (S36)

This indicates that this pair of EPs comes from infinity. To illustrate the properties of this case, we plot eigen-frequencies, phase rigidity and evolution of eigenstates in Figs. S10, and S11.

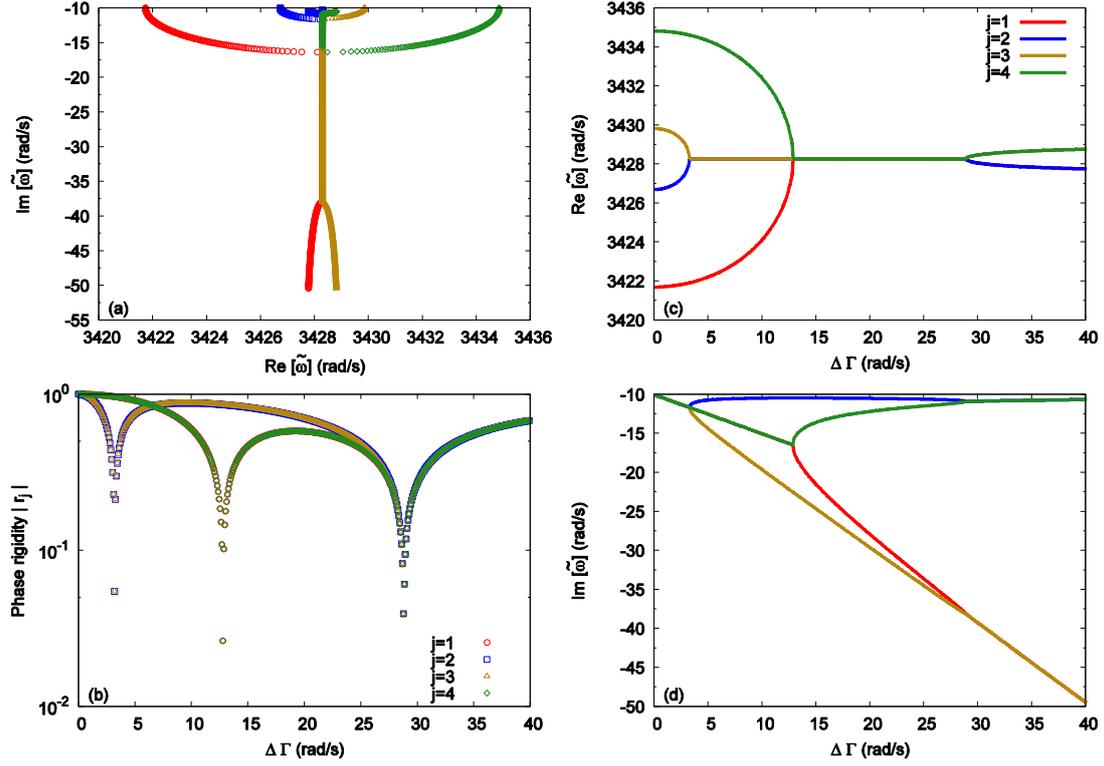

**Figure S10.** (a) Trajectory of eigen-frequency $\tilde{\omega}_j$ in the complex frequency plane. (b) Norm of phase rigidity $|r_j|$ for each state as function of loss difference $\Delta\Gamma$. (c) Real parts and (d) imaginary parts of eigen-frequency $\tilde{\omega}_j$ as function of loss difference $\Delta\Gamma$. Parameters used here are $\omega_1 = 3427.59\text{Hz}$, $\omega_2 = 3429.00\text{Hz}$, $\kappa = -2.5\text{Hz}$, $t = -4.0\text{Hz}$, and $\Gamma_0 = 10\text{Hz}$.

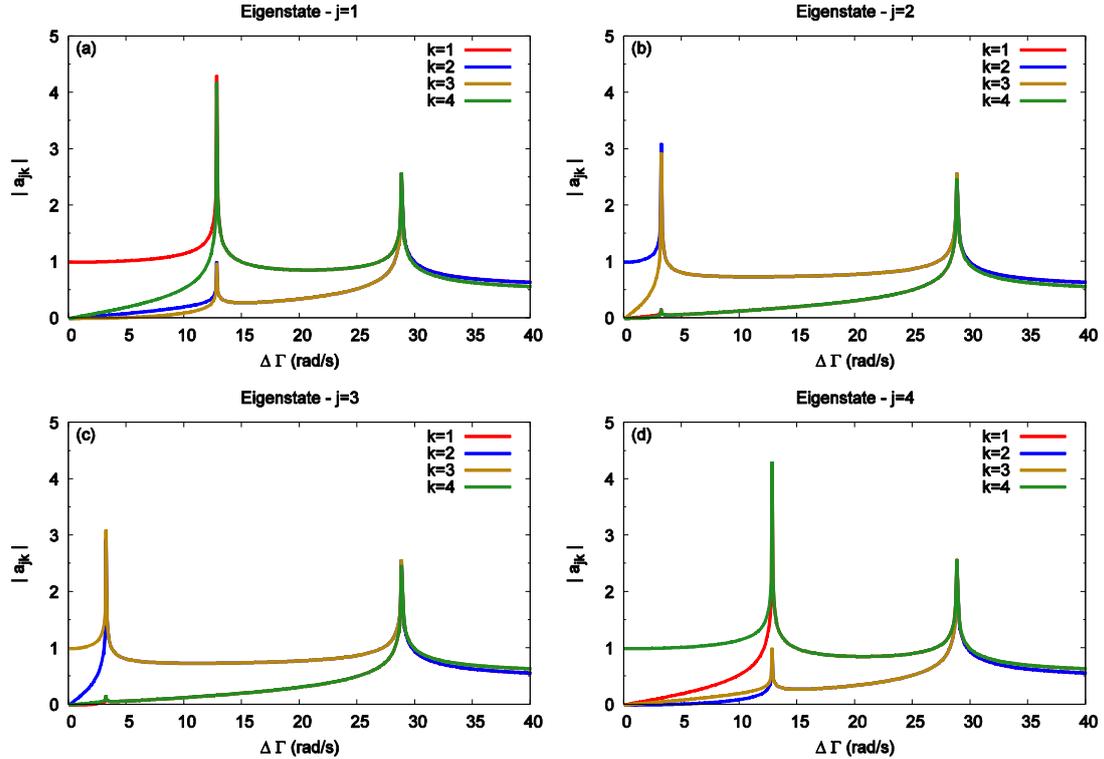

**Figure S11.** Evolution of eigenstate-$j$ in Figure S10 as function of loss difference $\Delta\Gamma$ in the $\Gamma_0$-representation, in which $a_{jk} = \langle \phi^L_{(0),k} | \phi^R_j (\Delta\Gamma) \rangle$, and $\{\langle \phi^L_{(0),k}|; k=1,2,3,4\}$ are eigenstates at $\Delta\Gamma = 0$.

## 7 Case four: the "kissing line"

We use the term "kissing line" to mean that two or more EPs collide in the complex parameter plane, and mathematical conditions are already given in Eq. (S15). To illustrate the properties of this case, we plot eigen-frequencies, phase rigidity and evolution of eigenstates in Figs. S12, and S13.

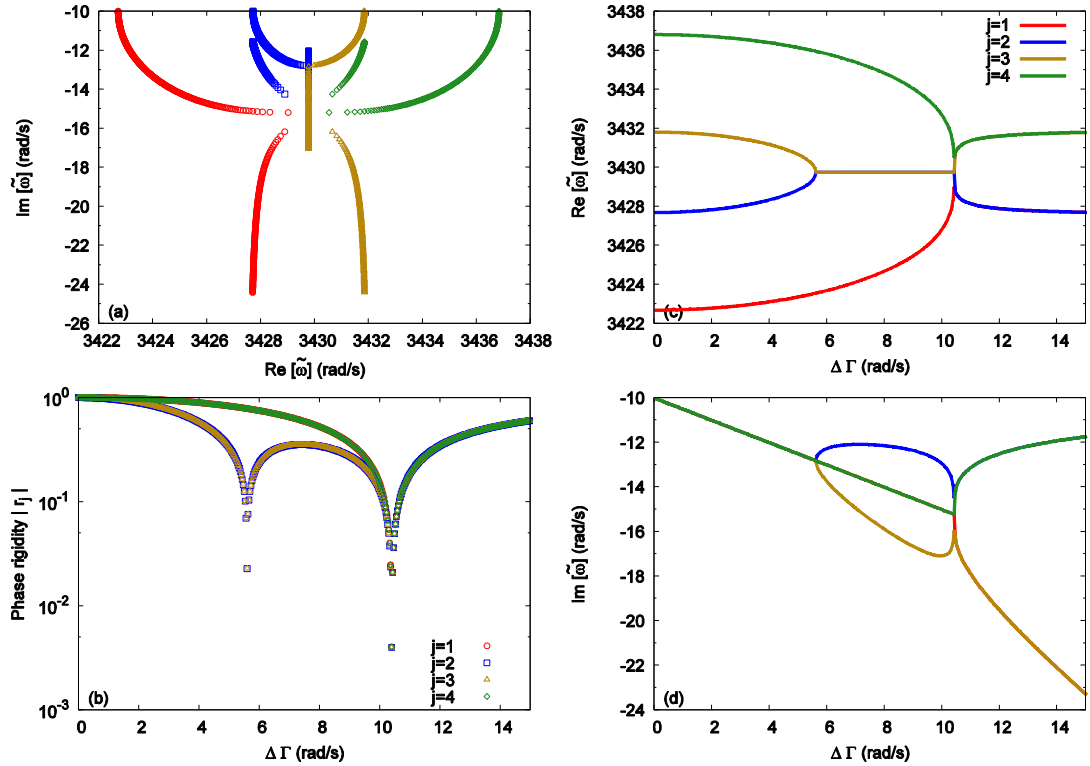

**Figure S12.** (a) Trajectory of eigen-frequency $\tilde{\omega}_j$ in the complex frequency plane. (b) Norm of phase rigidity $|r_j|$ for each state as function of loss difference $\Delta\Gamma$. (c) Real parts and (d) imaginary parts of eigen-frequency $\tilde{\omega}_j$ as function of loss difference $\Delta\Gamma$. Parameters used here are $\omega_1 = 3427.59\text{Hz}$, $\omega_2 = 3431.974631558\text{Hz}$, $\kappa = -2.5\text{Hz}$, $t = -4.0\text{Hz}$, and $\Gamma_0 = 10\text{Hz}$.

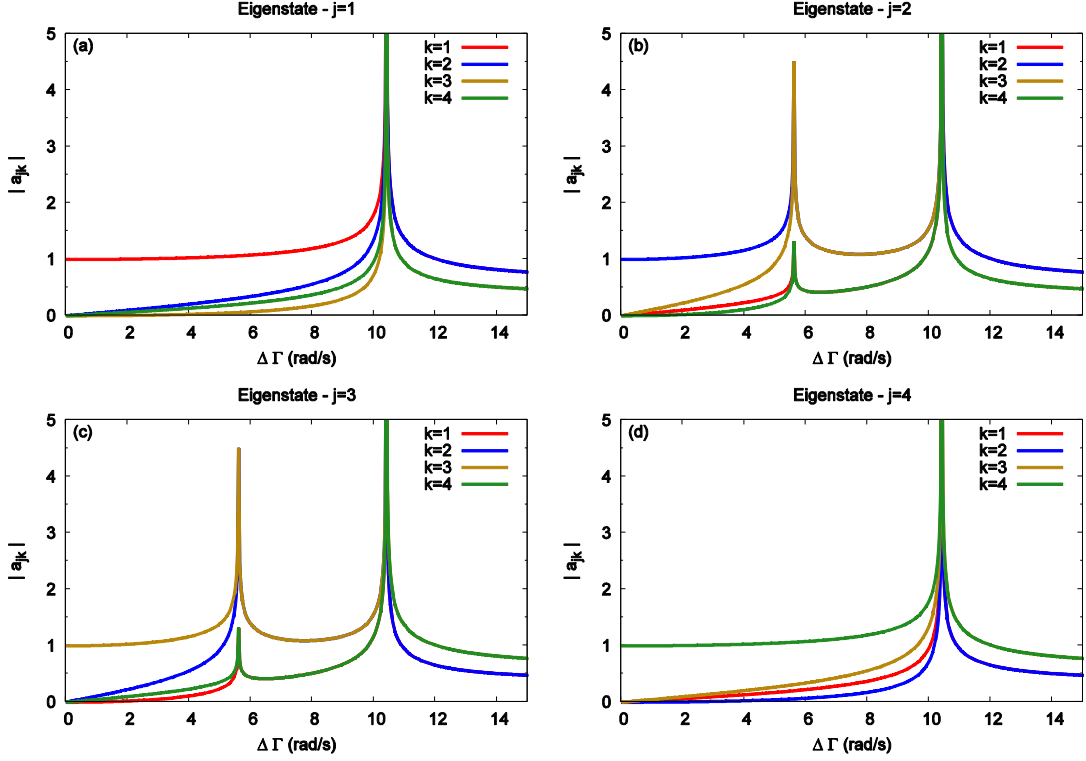

**Figure S13.** Evolution of eigenstate- $j$ in Figure S12 as function of loss difference $\Delta\Gamma$ in the $\Gamma_0$-representation, in which $a_{jk} = \langle \phi^L_{(0),k} | \phi^R_j(\Delta\Gamma)\rangle$, and $\{\langle \phi^L_{(0),k}|; k=1,2,3,4\}$ are eigenstates at $\Delta\Gamma = 0$.

## 8 Geometric phase around EPs

Since there is only one independent eigenvector at the EP, the chirality of EP could be defined analogous to the polarization of electromagnetic waves, namely that the chirality of EP is "right" if the eigenvector is $(-i,1)/\sqrt{2}$, and "left" if the vector at EP is $(i,1)/\sqrt{2}$. And it is known that for a single loop around an EP, the two eignfunctions will switch up to a sign which depends on the direction of the loop[2]. For example

$$\begin{pmatrix}\phi_1\\ \phi_2\end{pmatrix} \to \begin{pmatrix}\phi_2\\ -\phi_1\end{pmatrix} \text{(conter-clockwise)}, \quad \begin{pmatrix}\phi_1\\ \phi_2\end{pmatrix} \to \begin{pmatrix}-\phi_2\\ \phi_1\end{pmatrix} \text{(clockwise)}, \quad (S37)$$

in which the arrow means the states loop around EP by one cycle. Then looping around single EP in the conter-clockwise direction for one loop will follow as

$$\frac{1}{\sqrt{2}}\begin{pmatrix}\pm i\\ 1\end{pmatrix} \xrightarrow{loop} \frac{1}{\sqrt{2}}\begin{pmatrix}1\\ \mp i\end{pmatrix} = \exp\left[i\frac{\mp \pi}{2}\right] \times \frac{1}{\sqrt{2}}\begin{pmatrix}\pm i\\ 1\end{pmatrix}. \quad (S38)$$

This means that encircling single EP two loops gives out geometric phase $\pm\pi$.

## 9 Topological difference between "a" and "b"

We show that the topological characteristics of the two CS-1 singularities are different in regions "*a*" and "*b*". Since these two singularities have opposite signs in Class II-a, a cyclic variation of $\Delta\Gamma$ encircling both EPs produces no geometric phase due to phase cancellation. This can be seen from Eq. (S24) by setting $t=0$. As $t$ is increased beyond the white dashed line, states 2 and 3 become inverted inducing the chirality of the first EP to change sign, whereas the chirality of the other EP remains unchanged. This creates a pair of CS-1 EPs of the same chirality in Class II-b. Thus, a cyclic variation of $\Delta\Gamma$ encircling both EPs produces a geometric phase $\pm\pi$. This is consistent with the singularity on the yellow solid line we have found before.